\documentclass[aps,preprint,showpacs,preprintnumbers,amsmath,amssymb,floats,groupedaddress]{revtex4-1} 
\usepackage{graphicx}
\usepackage[english]{babel}
\usepackage{amsmath}
\usepackage{amssymb}
\usepackage{color}

\newcommand{\be}{\begin{eqnarray}} 
\newcommand{\ee}{\end{eqnarray}}

\begin{document}

\title{Quantum-critical resistivity of strange metals in a magnetic field} 

\author{Chandra M. Varma}
\affiliation{\footnote{Recalled Professor}Physics Department, University of California, Berkeley, Ca. 94704\\
\footnote{Emeritus Distinguished Professor}Physics Department, University of California, Riverside, Ca. 92521}

\begin{abstract}
Resistivity in the quantum-critical fluctuation region of several metallic compounds such as the cuprates, the heavy-fermions, Fe-chalogenides and pnictides, Moir\'e bi-layer graphene (MBLG) and WSe$_2$,  is linear in temperature $T$ as well as in the magnetic field $H_z$ perpendicular to the planes.  Scattering of fermions by  the fluctuations of a time-reversal odd polar vector field ${\bf \Omega}$ 
 has been shown to give a linear in T resistivity and other marginal Fermi liquid properties.  An extension of this theory to an applied magnetic field is presented. Magnetic field is shown to generate a density of vortices in the field ${\bf \Omega}$ proportional to $H_z$. The elastic scattering of fermions from the vortices gives a resistivity linear in $H_z$ with the coefficient  varying as the marginal fermi-liquid susceptibility $\ln(\frac{\omega_c}{T})$. Quantitative comparison with experiments is presented for cuprates and MBLG .
\end{abstract}
\maketitle

High temperature cuprates \cite{Ginzburg-rev} \cite{CMV_RMP(2020)} have a linear in $T$ resistivity for doping in the region above T$_c$ which is bounded by a
 phase with a "pseudo-gap"  on one side and cross-over to a Fermi-liquid on the other. This and related anomalies \cite{CMVlosalamos1991} in this 
  region suggested a quantum-critical origin for the anomalies \cite{cmv1997} \cite{simon-cmv} and the prediction that the pseudogap phase breaks time-reversal and inversion symmetries. Linear in T resistivity 
  and other anomalies, similar to those in the cuprates, are also found in several Fe-based compounds in the
   fluctuation region of  their antiferromagnetic (AFM) quantum-critical point \cite{Hosono-rev, ShibauchiQCP, MGKim2015, ThermopowerFeAS}, 
  in heavy fermion compounds \cite{HvLRMP2007, PaschenYBRh}, as well as more recently in twisted bi-layer graphene \cite{Herrero2019, Efetov2019} and in 
  the twisted bi-layer compound WSe$_2$ \cite{Pasupathy2021} with hitherto un-discovered order parameters. 
An important recent discovery \cite{Analytis1} \cite{Shekhter2018} \cite{Efetov2021} \cite{Pasupathy2021} is that in all of them the resistivity  
 is linear also in an applied magnetic field $|H|$.  The magnitude of the magneto-resistivity is similar to the zero-field resistivity at temperature $T$ for $\mu_BH$ of $O(k_BT)$. Where investigated \cite{Hayes2018}, \footnote {G. Boebinger and A. Shekhter - private communication 2020} the linear in $|H|$ resistivity is found only for the component $H_z$ applied perpendicular to the planes. 
 %Linear in $T$ and in $H$ resistivity is also found \cite{ChaoYang2021} in a region of the phase diagram of a 2d superconductor- cooper pair insulator- anomalous metal system \cite{Kapitulnik_SIM_RMP}.
 
 Three important general points should be noted:  First, a transport scattering rate linear in $|H|$ and (nearly) independent of temperature can only be due to elastic  scattering of fermions from time-reversal odd axial objects
 induced by the magnetic field.  Second, the fact that only the component of the field orthogonal to the high conducting plane in all these metals is effective excludes magnetic moment due to spins in favor of magnetic moments due to orbital loop currents. (The inevitable coupling to spins produces a magneto-resistance $\propto H^2$, which from measurements away from criticality is estimated at 50 Tesla to be  few percent of the experimental value \cite{Shekhter2018}.) Third, the magnitude mentioned above implies that the theory of the linear in $|H|$ resistivity must be closely related to the theory which gives linear in $T$ resistivity. 
 
 A  theory that gives
 linear in $T$ resistivity and other anomalies in cuprates rests on 
 the theory of quantum critical fluctuations \cite{Aji-V-qcf1, Aji-V-qcf3} which are
prelude to a state of loop current order. The new experiments invite extension of this theory to the effects of a magnetic field. The occurrence of the linear in $T$ and in $H$ resistivity as well as the associated $T ~\ln (\omega_c/\pi T)$ entropy in the quantum-critical regions in at least all the other compounds where results are available \cite{HvLRMP2007, ShibauchiQCP} is to be expected if their
quantum-criticality is described by a model which maps to the quantum-xy model coupled to fermions (QXY-F). The mapping has been shown \cite{CMV-IOP-REV} for the planar ferro or antiferro-magnetic model or an incommensurate Ising model. Here, I will first present a theory for the magnetic field dependence of the resistivity in the cuprate compounds for which more quantitative information is available than the other compounds and briefly comment on the other cases. 

\begin{figure}[h]
    \includegraphics[width= 1.0\columnwidth]{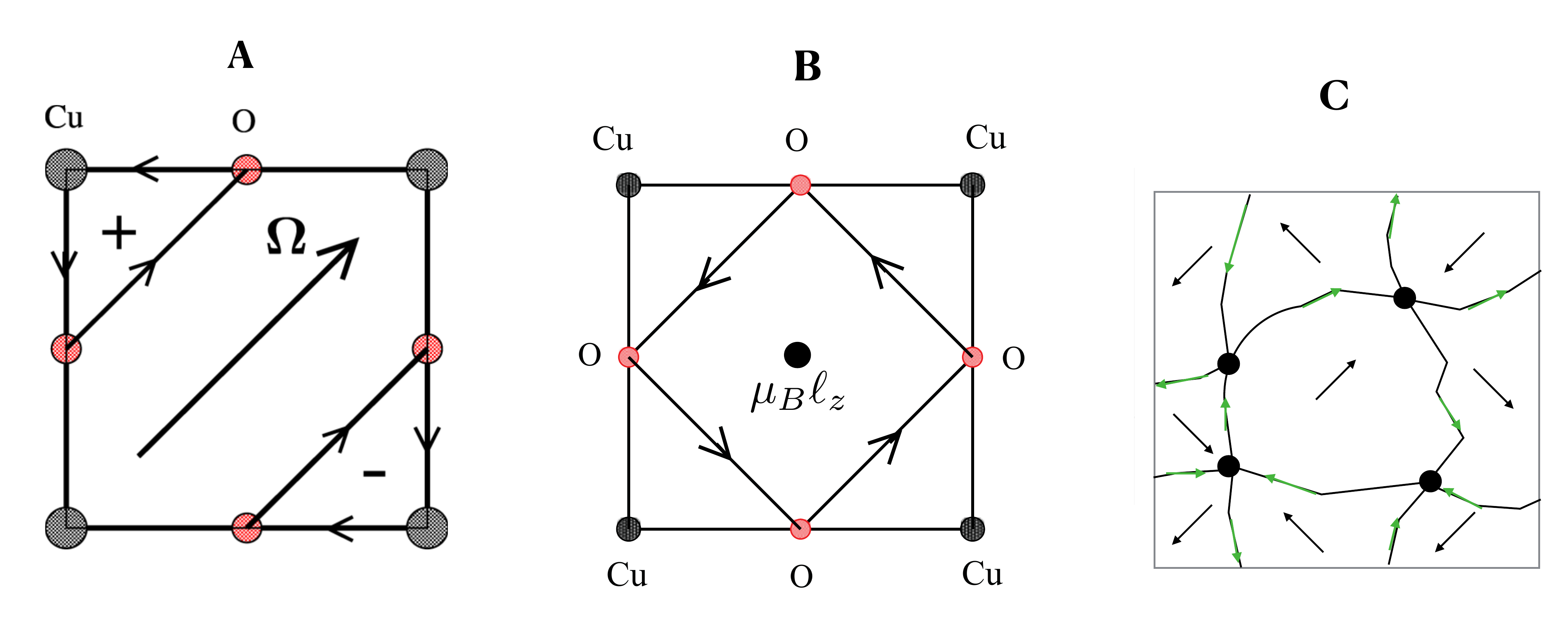}
  \caption{Representation of the current distribution in the cu-o unit cell for, $~$ A: the vector field ${\bf \Omega}$ which has one of four possible angles $\theta_i$ in the unit-cell $i$, $~$B: the angular momentum ${\bf \ell}_z$ which is a generator of rotations of ${\bf \Omega}$ and has a magnetization at its core. $~$ C:  Represents a fluctuation of ${\bf \Omega}$ over regions of many cells. A current represented by the green arrows runs at the boundary between any two orientations of ${\bf \Omega}$. At all corners of the variations in ${\bf \Omega}$ a vortex or ${\bf \ell}_z$, represented by the black dot, is required to exist. At $H=0$, the vortices are of equally up and down orientations. But an applied finite $H$ leads to  a net orbital angular momentum due to unequal density of vortices of different orientation.}
 \label{FigLzOm}
 \end{figure}

Loop-current order in cuprates can be represented as a time-reversal odd polar vector ${\bf \Omega}$ on a lattice, sketched in Fig. (\ref{FigLzOm} - A).  Using conservation laws alone  Else and Senthil \cite{ElseSenthil2021} have recently  shown that to get resistivity proportional to $T$ for $T \to 0$ in the pure limit, the critical fluctuations must be of an order parameter of such a symmetry. Such an  order parameter has indeed been found to be consistent with experiments using a variety of different techniques \cite{Kaminski-diARPES, Bourges-rev, Hsieh2017, Armitage-Biref, Shu2018}.

 Consider the orbital magnetic susceptibility of the model in its quantum-critical regime. This is obtained from the  critical fluctuations already derived  in Refs. 
   \cite{Aji-V-qcf1,  ZhuChenCMV2015, ZhuHouV2016, Hou-CMV-RG}. The model at $H = 0$ is specified by the interaction energy of the angles $\theta_{i, \tau}$ of ${\bf \Omega}$ at neighboring sites, by 
 the kinetic energy due to their angular momentum $(\partial \theta_{i \tau}/\partial \tau)$,  
 and  the coupling of  spatial and temporal fluctuations in $\theta_{i, \tau}$ to the fermions. The action is \cite{Aji-V-qcf1}
\be
\label{qxy}
S &=& \int_0^{\beta} d\tau ~J \sum_{i<j}\cos\Big({\bf \theta}_{i,\tau} - {\bf \theta}_{j,\tau}\Big) + C^{-1} \sum_i\Big(\frac{d {\bf \theta}_i}{d\tau}\Big)^2 
+ S_{CF}. \\
S_{CF} &=&  \int_0^{\beta} d\tau \sum_i  i \alpha~\nabla \theta_{i,\tau} \cdot \psi^+{\bf j}_{i,\tau} \psi  
%+ \alpha'  ~\frac{d {\bf \theta}_i}{d\tau}  \psi^+({\bf r}\times {\bf p})_{i,\tau} \psi.
\ee
   $S_{CF}$ is the coupling of the fluctuations to the fermions  \cite{CMV-IOP-REV} through the coupling of collective mode current $\propto i\nabla \theta$ to the fermions current  operator ${\bf j}$. $\psi, \psi^+$ are the fermion field operators.  
  We have neglected a term
  for the four-fold anisotropy of $\theta$ on the lattice because it has been shown to be irrelevant in the quantum-critical region \cite{Aji-V-qcf1}.

   The QXY-F model, just as the classical XY model, does not belong to the universality class of the Ginzburg-Landau-Wilson
 theories and their quantum extensions. The quantum-critical fluctuations are driven by proliferation of topological defects, 2D spatial vortices, 
  and warps which are spatially local events interacting logarithmically in imaginary time  \cite{Aji-V-qcf1}. 
  The critical correlations of (\ref{qxy}) after integrating over the fermions, 
  $C({\bf r}, \tau) \equiv <e^{-i \theta({\bf r}, \tau)} e^{i \theta(0, 0)}>$, 
  have been obtained by quantum-monte-carlo calculations \cite{ZhuChenCMV2015, ZhuHouV2016}
as well as derived by 
renormalization group \cite{Hou-CMV-RG}.  It is shown in an Appendix in Ref. \cite{ASV2010} that the orbital magnetic susceptibility $\chi_{LL}({\bf r}, \tau)$ defined by (\ref{chi}),  are proportional to those of $C({\bf r}, \tau)$.
Near criticality the  (dimensionless) dynamic orbital magnetic susceptibility is
 \be
 \label{chi}
\chi_{LL}({\bf r}, \tau) \equiv \mu_B^2 <{\bf L}^+_z({\bf r}, \tau){\bf L}_z({\bf 0}, 0)> =  \mu_B^2 <L_z^2> \frac{\tau_c^2}{\tau} e^{-(\tau/\xi_{\tau})^{1/2}} 
\ln \frac{r}{a} e^{-r/\xi_r}.
\ee
$\mu_B^2<L_z^2>$ is the expectation value of the square of the magnitude of the orbital magnetic moment per unit-cell volume. We take this to be given by the amplitude of the measured \cite{Bourges-rev} ordered staggered moment per unit-cell $(\ell_z \mu_B)^2$. Since in the 2 d xy model, the dynamics are purely due to the phase fluctuations, the amplitude factor $<L_z^2>$ is nearly temperature independent in the region of interest. $\tau_c$ is the short time cut-off obtainable from experiments
and has a value of  $O(10^3 K)^{-1}$ \cite{ASV2010, CMV_RMP(2020)}, $a$ is the lattice constant.
  An important result that the spatial correlation length normalized to the lattice constant
 $\xi_r/a \approx log (\xi_\tau/\tau_c)$, the normalized temporal correlation. So the dynamical critical exponent $z = \infty$ and the problem is effectively local in space.
 The spectral function (\ref{chi})  is of the form  proposed 
   phenomenologically \cite{CMV-MFL} to give the fluctuations of marginal Fermi-liquid, rather than the $1/\tau^2$ of the Landau Fermi-liquid.   In terms of the frequency $\omega$ and temperature $T$, 
   \be
\label{chiom}
\chi_{LL}(\omega,T) =  \frac{\mu_B^2  <L_z^2>}{\omega_c}  \Big(\ln  \big|\frac{\omega_c}{max(\omega, \pi T)}\big| - i \tanh \frac{\omega}{2T}\Big),
\ee
at criticality.  $\omega_c = 1 /\tau_c$ is the ultra-violet cut-off. This functional form is also the principal result of theories on interesting models of  mathematical interest such as the SYK model \cite{SYK2021}, holographic models \cite{HongLiu2011}  and of other models \cite{Dumeitrescu2021}, \cite{CMVlosalamos1991,  Ruckenstein1991, BergHartnoll2020}.  
 
 The Hamiltonian for the coupling of a magnetic field couples to the angular momentum is
   \be
\label{SB}
H_B = - \mu_B \sum_i~{\bf H \cdot L}_{iz}(\tau),  
\ee
where ${\bf L}_{iz} \equiv i (\partial/\partial \theta_i)$. 
%If resistivity, regarded as a critical property is  linear both in $|H_z|$ and in $T$, the scaling dimension  of the magnetic %field 
%must be the same as that of $T$. I show this first and then  derive explicit results and present a quantitative comparison %of the theory and experiments.
  %From (\ref{SB}), it follows that the scaling dimension of $H_z$ are those of energy or temperature $T$ divided by those %of $ \mu_B L_z$. Since $\tau_c/\tau = e^{log (\tau_c/\tau)}$,  the correlation function of $ \mu_B {\bf L}_z$ is dimension-%less on scaling of $\tau$ or $T$ at criticality. 
% So the dimension of $H$ is the same as that of $T$.  As usual such an argument is uncertain to logarithmic corrections, %which will be found in the microscopic theory below. 
 In the quantum-critical regime $H_z$ induces a static macroscopic $<L_z>$ given by
 \be
 \label{statL}
\mu_B  <L_z> =  \chi'_{LL} H_z,~~\chi'_{LL}(T) =  \frac{\mu_B^2 \ell_z^2}{\omega_c} \log \big(\frac{\omega_c}{\pi T}\big).
 \ee
From the experimental observations \cite{Bourges-rev} that the ordered staggered moment per-cell is about $0.1 \mu_B$ , and $\omega_c \approx 2000 K$ \cite{ASV2010, CMV_RMP(2020)}, $\chi'_{LL}$ is estimated to be about $10^{-5}\mu_B^2/(Kelvin-cell)$. So a magnetic field of 50 Tesla  can be estimated to  produce a static magnetization $ \approx  5 \times 10^{-4}\mu_B$'s, not including the numerical factor due to the logarithmic temperature dependence in $\chi_{LL}$. An important question in the present context is how such a moment would be distributed. To think of this, it is useful to know the physical description of  $\ell_z$, the quasi-quantized unit of orbital angular momentum in the present problem. A loop-current carrying the lattice representation of  angular momentum is shown in Fig. (\ref{FigLzOm} - B) \cite{SunFradkin2008}.
It has been shown \cite{ASV2010, He-V-neutrons, HeMooreV2012} to be  the generator of rotations of the 
 magneto-electric vector ${\bf \Omega}$ in the plane, from one of its four orientations to the clockwise or anti-clockwise orientation:
 \be
 e^{i (\pi/2) {\bf \ell}_z} |\hat{{\bf \Omega}}> = |\hat{\bf{\Omega}} + \pi/2>.
 \ee
 The pictorial representation in Fig. (\ref{FigLzOm} - c) of
 $L_z$  corresponds to a vortex in the vector field ${\bf \Omega}$ with quantized angle but a magnetic moment given by the
 area and the current carried by the core cell around which the four-orientations of ${\bf \Omega}$ meet. Over long wave-lengths, one may ignore the granularity of the lattice so that $L_z$ is similar to the vortex
 in more familiar $U(1)$ fields such as superconductors in a magnetic field or superfluids in rotation. Instead of quantization of the  magnetic moment in terms of fundamental constants, it is non-universal and given by the magnitude of the vectors $\Omega$ which have very weak temperature dependence. 
 The magnitude of an individual moment $\ell_z$  is similar to the measured value of ${\bf \Omega}$ in the ordered state of about $0.1 \mu_B$ \cite{Bourges-rev}. Therefore the density of such moments $n_L$ is about $5 \times 10^{-3}$/unit-cell for a field of 50 Tesla so that their separation is about 50 unit-cells. In an ordered state of ${\bf \Omega}$, such moments would crystallize at low enough temperature due to their long-range mutual interactions. But we are considering the region in which they live in a bath of ${\bf \Omega}$'s quantum-fluctuating in time and space.  Therefore such moments would remain disordered at the temperatures of interest and diffuse at a very slow rate because of their enormous effective mass. If the motion of $<L_z>$  is very slow compared to the motion of fermions with 
 which they scatter, the scattering should be considered elastic. 
 
 \begin{figure}[h]
 \includegraphics[width= 0.8\columnwidth]{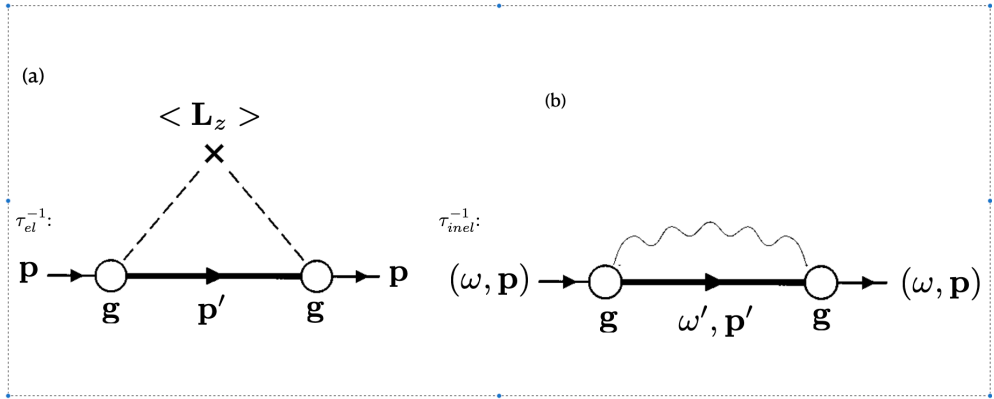}
\caption{(a) Elastic scattering  of fermions by vortices of angular momentum $<L_z>$, (b) Inelastic scattering of fermions by fluctuations $\chi"(\omega, T, {\bf q})$.}
 \label{Fig:Scatter}
\end{figure}

 The s-wave scattering rate $1/\tau_L$ of such local magnetic field generated defects can be easily estimated, See Fig. (\ref{Fig:Scatter}-a).
 \be
 \label{tauL}
 1/\tau_L &=& 2 \pi ~n_L (g_0 \mu_B \ell_z)^2~N(0), \\ 
 n_L & = & \frac{\chi'_{LL} H_z}{\mu_B \ell_z}~ \approx ~ \ell_z \frac{\mu_B H_z}{\omega_c} \ln(\omega_c/ \pi T).
 \ee
 $N(0)$ is the density of states of fermions at the chemical potential and $g_0$ is the coupling energy \cite{ASV2010} of the fermions to a vortex with orbital moment $\mu_B \ell_z$. 
 This is to be compared with the inelastic scattering of fermions by the fluctuations $\chi"(\omega,T)$, See Fig. (\ref{Fig:Scatter}-b). This is calculated from the analytic continuation of the imaginary part of the self-energy at zero-frequency which has been often derived, \cite{CMV-MFL, Pelzer1991, CMV_Lorentz}
  \be
  \label{Sig}
  1/\tau(T) &= & 2 U Im \Sigma"(0,T), \\
 \Sigma (i \omega_n) &=& g_0^2 \sum_{\omega_m, \bf k} G({\bf k}, i\omega_m) \chi (i \omega_n - i \omega_m),
 \ee 
 $U$ is a dimensionless Umklapp factor, which is necessary for finite resistivity.  Recently, in an asymptotically exact theory for resistivity due to fluctuations of the QXY-CF model, it has been shown that $U$ is temperature independent \footnote{H. Maebashi, C.M. Varma - unpublished}. A way to estimate $U$ is to compare the transport scattering rate with the imaginary part of the self-energy in the direction on the Fermi-surface of maximum velocity. This gives $U$ of $O(1)$ \cite{CMV_RMP(2020)} for the cuprates where both have been measured. Eq. (\ref{Sig}) gives
 \be
 \label{tauT}
 1 /\tau(T) \approx &~ \pi U ~ (g_0 \mu_B \ell_z)^2 ~N(0) \frac{k_BT}{\omega_c}.
 \ee
  $1/\tau_L$  and $1/\tau(T)$ are of similar magnitude at $\mu_B H/k_BT$ of $O(1)$ for $\ln (\omega_c/\pi T) \approx 1$. They are similar because   the inelastic scattering rate comes from the imaginary part of the same fluctuations whose real part gives $n_L$ to give the elastic scattering rate and the coupling energy to fermions is identical. 
 Specifically the ratio of the scattering rates is
 \be
 \label{predic}
 (1/\tau_L)\div (1/\tau(T)) \approx \frac{2 \ell_z}{U} \frac{\mu_B H}{k_BT} \ln(\omega_c/ \pi T).
 \ee
\begin{figure}[h]
 \includegraphics[width= 0.8\columnwidth]{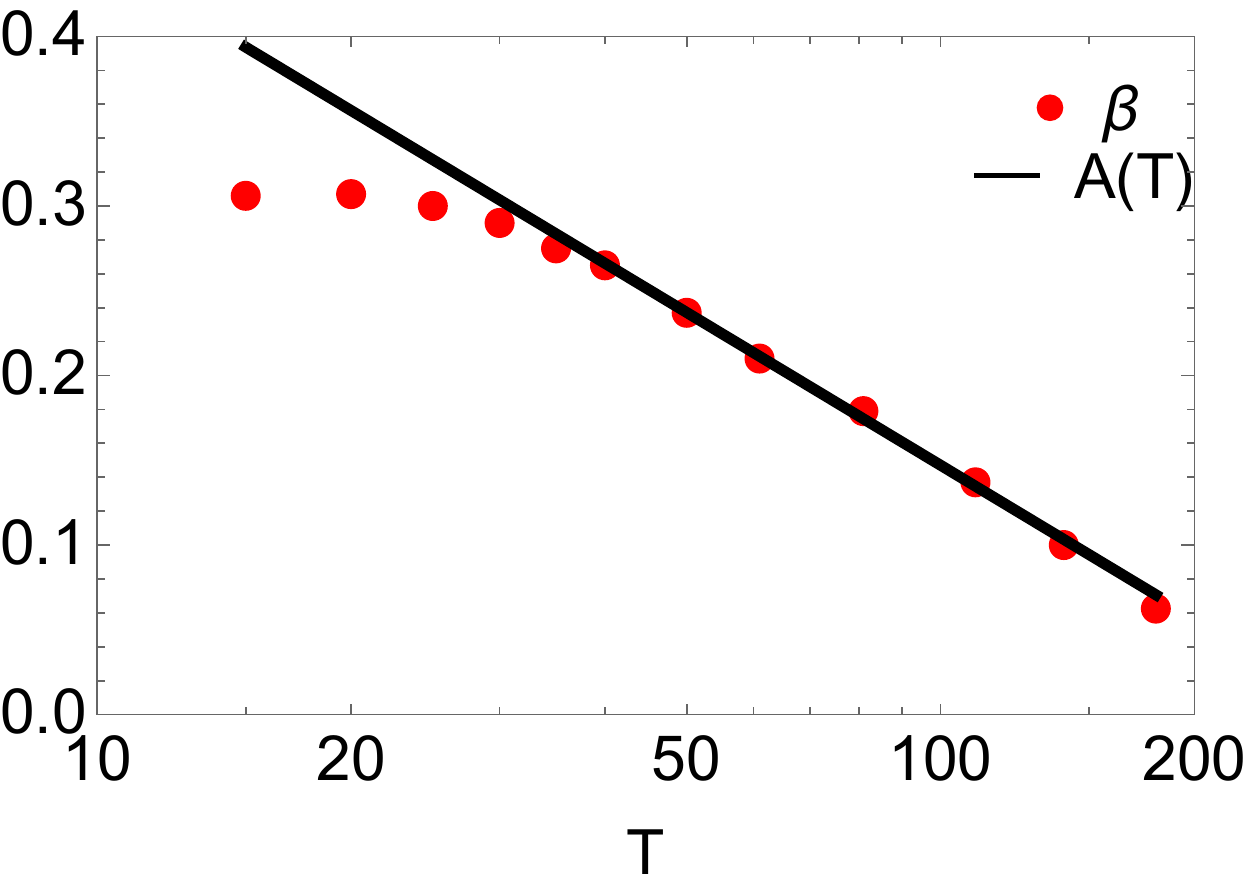}
\caption{The temperature dependence of the linear in H resistivity in $La_{2-x}Sr_xCuO_4$, for $x=0.19$. The data shown as red dots is taken from Fig. 1 b of Ref. (\cite{Shekhter2018}). $\beta(T)$ is obtained from the fit to the resistivity (after subtracting a small residual value) $\rho(T,H) -\rho_0 = \alpha ~ T + \beta(T) ~H$ at $H$ = 70  Tesla. $A(T) = 0.14 \ln\big|\frac{\omega_c}{\pi T}\big|$, with $\omega_c \approx 1600 K$.}
 \label{Fig:rho/B}
\end{figure}
The result (\ref{predic}) is subject to a cut-off at low temperatures if one is not at critical parameters, (the critical point is  also expected to shift in a magnetic field if the usual magnetic susceptibility of the system is different on the two sides of critical point) and a high temperature cut-off on the scale of the upper cut-off $\omega_c$. 

We can compare the result in Eq. (\ref{predic}) quantitatively with experiments. The data for the resistivity in the most extensively investigated case, for a cuprate near criticality, is represented in Ref. \cite{Shekhter2018} by $\rho(T,H) = \alpha k_BT + \beta(T) \mu_B H$. We can write using Eqs. (\ref{tauT}) and (\ref{predic}) that  $\beta(T) = \alpha \frac{2 \ell_z}{U} \ln(\omega_c/ \pi T)$.  $\beta(T)$ from low $T$ to the highest available temperature, $180 K$, and a logarithmic fit to it by $0.14 \ln (1500/\pi T)$ are given in Fig. (\ref{Fig:rho/B}). The coefficient $0.14$ should be compared with $0.19$ that is estimated from parameters above and the value of $\alpha \approx 1.1$ deduced in the experiment \cite{Shekhter2018}.  A logarithmic fit appears reasonable for $T \gtrsim 30 K$ below which the data saturates. The parameter $\omega_c$ is about $1600 K$, which may be compared with the $O(3000) K$ deduced \cite{CMV_RMP(2020)} from the fit to the logarithmic $C_v/T$ measured \cite{Tailleferspht} between $0.3 K$ and $10 K$. The peril of deducing a number from a logarithm in a range far above the data should be kept in mind. 
The data in Fig. 1a in \cite{Shekhter2018} shows systematic rounding towards zero below about 30 K even in a field of $70$ Tesla. One may be tempted to ascribe it to not being very close to criticality, but a closer look at all the data at various fields suggests  a more mundane reason. The data shows a large region of rounding from the zero-field transition temperature ($\approx 41 K$) towards zero resistivity at low temperatures even in large fields.  This is generally the rule in 2d strongly type II superconductors or superconducting films due to an enhanced region of phase fluctuations in a field.  

An independent way to test the prediction made here is to see if a direct measurement of  magnetization in the range in which the resistivity satisfies Eqn. ({\ref{predic}) shows  the same logarithmic enhancement.

 I now briefly discuss the other compounds, to begin with those for which the quantum-criticality is that of antiferromagnetism.
 Significantly, the important critical fluctuations for planar and incommensurate Ising ferromagnets or anti-ferromagnets (or charge density waves) are of the phase variable given by the xy model \cite{McMillanCDW, CMV-IOP-REV}. It is very interesting to note  that the  measured spectral functions for 
 the quantum-critical fluctuations for planar antiferromagnetism in $BaFe_{1.85} Co_{0.15 }As_2$ \cite{Inosov2010} or incommensurate anti-ferromagnetism in the heavy fermion $CeCu_6$ \cite{Schroder1, Schroder2} are consistent with the product form in momentum and energy \cite{SchroderZhuV2015} as in Eq. (\ref{chi}) for the QXY-SF model. 
 
The data \cite{Analytis1} in $BaFe_2(As_{1-x}P_x)_2$ with $T_c \approx 30 K$, is available only to 60 K with fields up to 59 Tesla has a severe rounding of resistivity towards zero at low temperatures for fields less than 50 Tesla so that  linearity of $H$ above this field is observed only in a narrow range of temperatures.  We therefore cannot 
usefully compare the data in $BaFe_2(As_{1-x}P_x)_2$. The fit of the resistivity data  made as 
 $\propto \sqrt{(\mu_BH)^2 + (k_BT)^2}$ made earlier  \cite{Analytis1} is not good under closer examination of the detailed data kindly received from the authors (I. Hayes - private communication - Dec. 2021). That fit also does not work for the cuprate or for the twisted bi-layer graphene \cite{Efetov2021}, as stated by the authors. However,
 an $H^2$ dependence of magneto-resistance at low fields is conventional and well understood and there is no reason why it should be completely absent in the metals
 under discussion.
 
 \begin{figure}[h]
 \includegraphics[width= 0.8\columnwidth]{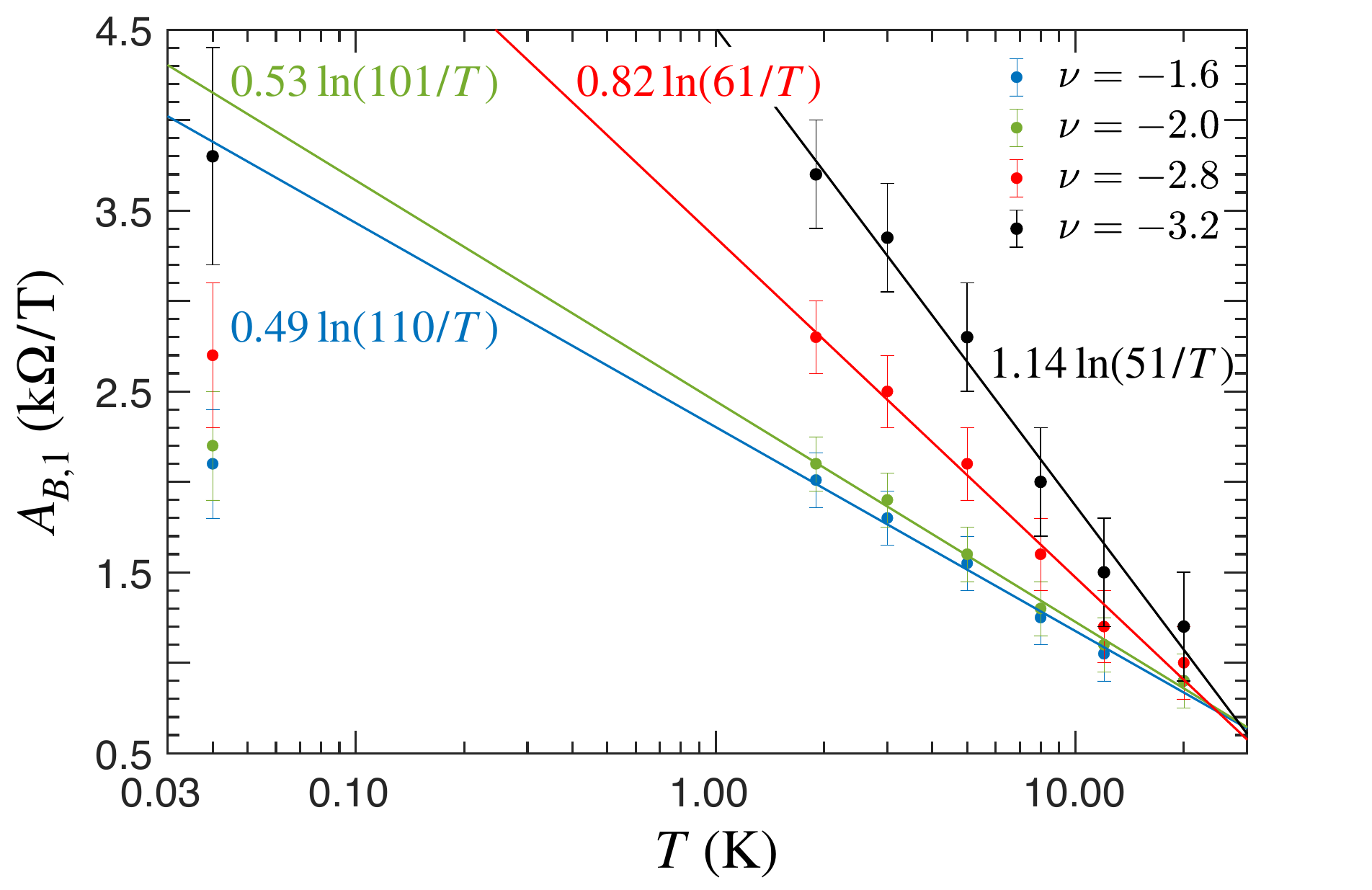}
\caption{The temperature dependence of the linear in H resistivity in twisted bi-layer graphene at the dopings where the resistivity at $H=0$ is most nearly linear in $T$ down to the lowest temperature measured. The data is from Ref. (\cite{Efetov2021}) and replotted by the authors of that paper and provided to me. $A_{B,1} \equiv \beta(T)$ is obtained just as for the cuprate compound in Ref. (\ref{Fig:rho/B}).}
 \label{Fig:rho/B-graphene}
\end{figure}

  The relevant order parameter for Twisted bi-layer (TB)-graphene and TB-$WSe_2$ is not known yet from experiments, although there are theoretical calculations suggestive of loop-current ordered states \cite{Zaletel2020,  Berg2021} in TB-graphene. TB-$WSe_2$ is similar except for the large spin-orbit coupling. Their structure has a triangular motif
 and it is expected that the nearest neighbor repulsion is comparable to the kinetic energy. In this situation, loop-current order is a likely instability \cite{Weber-Giam-V, Nand_levitov_BG, Lijun-graphene}. It should be ascertained if only the component  of the magnetic field perpendicular to the plane is responsible for the resistance linear in the field. If this holds,  more experiments to test the time-reversal, inversion and possible chirality given by loop-currents are suggested to decipher their long-range order. The data on $WSe_2$ is not yet detailed enough to compare with theory for $\beta(T)$, but it is for TB-graphene \cite{Efetov2021}. This is plotted in (\ref{Fig:rho/B-graphene}). The logarithmic fit is found with a coefficient varying from about 0.5 to about 1 and an upper cut-off $\omega_c$ varying from about 50 K to about 100 K. There are no independent numbers from other experiments to compare. But the scale of the fluctuation energies an order of magnitude smaller than the cuprates appears reasonable and the magnitude suggests that the magnitude of the orbital moment in the ordered state is similar to that in the cuprates. The saturation at the lowest point at $40 mK$ is almost certainly due to rounding of resistivity due to impending superconductivity. 
  
 The experiments in a magnetic field test a crucial microscopic aspect underlying the applicability of the theory of quantum-fluctuations of the xy-model. They identify that the xy model pertains to a vector which is odd in time-reversal and inversion. Only then can the magnetic field couple to the magnetic moment which is a generator of rotations of the vector characterizing in-plane order.  It is already understood that d-wave superconductivity is not possible if the self-energy of the fermions is angle-independent as it is in cuprates without the
 fermions coupling to the fluctuations of angular momentum \cite{ASV2010, CMV_RMP(2020)}. This is consistent with the information about the physics of d-wave pairing in cuprates obtained from angle-resolved photoemission \cite{Bok_ScienceADV}.
 To conclude, one might also add that the mechanism of superconductivity in all these systems is inevitably related
 to the fluctuations which give resistivity linear in $T$ and in $H$.
 
 {\it Acknowledgements}: I have benefitted enormously from discussions with Arkady Shekhter both of the experimental data and the theory. Discussions of the theory with Lijun Zhu and Vivek Aji have been invaluable. I have also benefitted by discussions of the state of experiments with Greg Boebinger on cuprates, James Analytis and Ian Hayes on the Fe-compound, with Augusto Ghiotto and Abhay Pasupathy on WSe$_2$, with A. Jaoui on MBLG, and with Erez Berg on the theory for TB-graphene.
This work was partially performed at the Aspen Center for Physics, which is supported by National Science Foundation grant PHY-1607611.

%\bibliography{REFjuly2019.bib}

\begin{thebibliography}{58}%
\makeatletter
\providecommand \@ifxundefined [1]{%
 \@ifx{#1\undefined}
}%
\providecommand \@ifnum [1]{%
 \ifnum #1\expandafter \@firstoftwo
 \else \expandafter \@secondoftwo
 \fi
}%
\providecommand \@ifx [1]{%
 \ifx #1\expandafter \@firstoftwo
 \else \expandafter \@secondoftwo
 \fi
}%
\providecommand \natexlab [1]{#1}%
\providecommand \enquote  [1]{``#1''}%
\providecommand \bibnamefont  [1]{#1}%
\providecommand \bibfnamefont [1]{#1}%
\providecommand \citenamefont [1]{#1}%
\providecommand \href@noop [0]{\@secondoftwo}%
\providecommand \href [0]{\begingroup \@sanitize@url \@href}%
\providecommand \@href[1]{\@@startlink{#1}\@@href}%
\providecommand \@@href[1]{\endgroup#1\@@endlink}%
\providecommand \@sanitize@url [0]{\catcode `\\12\catcode `\$12\catcode
  `\&12\catcode `\#12\catcode `\^12\catcode `\_12\catcode `\%12\relax}%
\providecommand \@@startlink[1]{}%
\providecommand \@@endlink[0]{}%
\providecommand \url  [0]{\begingroup\@sanitize@url \@url }%
\providecommand \@url [1]{\endgroup\@href {#1}{\urlprefix }}%
\providecommand \urlprefix  [0]{URL }%
\providecommand \Eprint [0]{\href }%
\providecommand \doibase [0]{http://dx.doi.org/}%
\providecommand \selectlanguage [0]{\@gobble}%
\providecommand \bibinfo  [0]{\@secondoftwo}%
\providecommand \bibfield  [0]{\@secondoftwo}%
\providecommand \translation [1]{[#1]}%
\providecommand \BibitemOpen [0]{}%
\providecommand \bibitemStop [0]{}%
\providecommand \bibitemNoStop [0]{.\EOS\space}%
\providecommand \EOS [0]{\spacefactor3000\relax}%
\providecommand \BibitemShut  [1]{\csname bibitem#1\endcsname}%
\let\auto@bib@innerbib\@empty
%</preamble>
\bibitem [{\citenamefont {Ginsburg}(1994)}]{Ginzburg-rev}%
  \BibitemOpen
  \bibinfo {editor} {\bibfnamefont {D.}~\bibnamefont {Ginsburg}},\ ed.,\
  \href@noop {} {\emph {\bibinfo {title} {Physical Properties of High
  Temperature Superconductors}}}\ (\bibinfo  {publisher} {World Scientific,
  Singapore},\ \bibinfo {year} {Vol I (1989), Vol II (1990), Vol III (1992),
  Vol. Iv (1994)})\BibitemShut {NoStop}%
\bibitem [{\citenamefont {Varma}(2020)}]{CMV_RMP(2020)}%
  \BibitemOpen
  \bibfield  {author} {\bibinfo {author} {\bibfnamefont {C.~M.}\ \bibnamefont
  {Varma}},\ }\href {\doibase 10.1103/RevModPhys.92.031001} {\bibfield
  {journal} {\bibinfo  {journal} {Rev. Mod. Phys.}\ }\textbf {\bibinfo {volume}
  {92}},\ \bibinfo {pages} {031001} (\bibinfo {year} {2020})}\BibitemShut
  {NoStop}%
\bibitem [{\citenamefont {Varma}(1994)}]{CMVlosalamos1991}%
  \BibitemOpen
  \bibfield  {author} {\bibinfo {author} {\bibfnamefont {C.}~\bibnamefont
  {Varma}},\ }\href@noop {} {\emph {\bibinfo {title} {"Theoretical framework
  for the normal state of copper oxide metal" in Strongly Correlated Electronic
  Systems, the Los Alamos Symposium 1991}}}\ (\bibinfo  {publisher}
  {Addison-Wesley, Reading MA, 1994},\ \bibinfo {year} {1994})\ pp.\ \bibinfo
  {pages} {573--603}\BibitemShut {NoStop}%
\bibitem [{\citenamefont {Varma}(1997)}]{cmv1997}%
  \BibitemOpen
  \bibfield  {author} {\bibinfo {author} {\bibfnamefont {C.~M.}\ \bibnamefont
  {Varma}},\ }\href {\doibase 10.1103/PhysRevB.55.14554} {\bibfield  {journal}
  {\bibinfo  {journal} {Phys. Rev. B}\ }\textbf {\bibinfo {volume} {55}},\
  \bibinfo {pages} {14554} (\bibinfo {year} {1997})}\BibitemShut {NoStop}%
\bibitem [{\citenamefont {Simon}\ and\ \citenamefont
  {Varma}(2002)}]{simon-cmv}%
  \BibitemOpen
  \bibfield  {author} {\bibinfo {author} {\bibfnamefont {M.~E.}\ \bibnamefont
  {Simon}}\ and\ \bibinfo {author} {\bibfnamefont {C.~M.}\ \bibnamefont
  {Varma}},\ }\href {\doibase 10.1103/PhysRevLett.89.247003} {\bibfield
  {journal} {\bibinfo  {journal} {Phys. Rev. Lett.}\ }\textbf {\bibinfo
  {volume} {89}},\ \bibinfo {pages} {247003} (\bibinfo {year}
  {2002})}\BibitemShut {NoStop}%
\bibitem [{\citenamefont {Ishida}\ \emph {et~al.}(2009)\citenamefont {Ishida},
  \citenamefont {Nakai},\ and\ \citenamefont {Hosono}}]{Hosono-rev}%
  \BibitemOpen
  \bibfield  {author} {\bibinfo {author} {\bibfnamefont {K.}~\bibnamefont
  {Ishida}}, \bibinfo {author} {\bibfnamefont {Y.}~\bibnamefont {Nakai}}, \
  and\ \bibinfo {author} {\bibfnamefont {H.}~\bibnamefont {Hosono}},\
  }\href@noop {} {\bibfield  {journal} {\bibinfo  {journal} {J. Phys. Soc. Jpn.
  78}\ }\textbf {\bibinfo {volume} {78}},\ \bibinfo {pages} {062001} (\bibinfo
  {year} {2009})}\BibitemShut {NoStop}%
\bibitem [{\citenamefont {Shibauchi}\ \emph {et~al.}(2014)\citenamefont
  {Shibauchi}, \citenamefont {Carrington},\ and\ \citenamefont
  {Matsuda}}]{ShibauchiQCP}%
  \BibitemOpen
  \bibfield  {author} {\bibinfo {author} {\bibfnamefont {T.}~\bibnamefont
  {Shibauchi}}, \bibinfo {author} {\bibfnamefont {A.}~\bibnamefont
  {Carrington}}, \ and\ \bibinfo {author} {\bibfnamefont {Y.}~\bibnamefont
  {Matsuda}},\ }\href {\doibase 10.1146/annurev-conmatphys-031113-133921}
  {\bibfield  {journal} {\bibinfo  {journal} {Annual Review of Condensed Matter
  Physics}\ }\textbf {\bibinfo {volume} {5}},\ \bibinfo {pages} {113} (\bibinfo
  {year} {2014})},\ \Eprint
  {http://arxiv.org/abs/https://doi.org/10.1146/annurev-conmatphys-031113-133921}
  {https://doi.org/10.1146/annurev-conmatphys-031113-133921} \BibitemShut
  {NoStop}%
\bibitem [{\citenamefont {Kim}\ \emph {et~al.}(2015)\citenamefont {Kim},
  \citenamefont {Wang}, \citenamefont {Tucker}, \citenamefont {Valdivia},
  \citenamefont {Abernathy},\ and\ \citenamefont {Chi}}]{MGKim2015}%
  \BibitemOpen
  \bibfield  {author} {\bibinfo {author} {\bibfnamefont {M.}~\bibnamefont
  {Kim}}, \bibinfo {author} {\bibfnamefont {M.}~\bibnamefont {Wang}}, \bibinfo
  {author} {\bibfnamefont {G.}~\bibnamefont {Tucker}}, \bibinfo {author}
  {\bibfnamefont {P.}~\bibnamefont {Valdivia}}, \bibinfo {author}
  {\bibfnamefont {D.}~\bibnamefont {Abernathy}}, \ and\ \bibinfo {author}
  {\bibfnamefont {e.~a.}\ \bibnamefont {Chi}, \bibfnamefont {Songxue}},\
  }\href@noop {} {\bibfield  {journal} {\bibinfo  {journal} {arXIV:1510.04167}\
  } (\bibinfo {year} {2015})}\BibitemShut {NoStop}%
\bibitem [{\citenamefont {Lv}\ \emph {et~al.}(2009)\citenamefont {Lv},
  \citenamefont {Gooch}, \citenamefont {Lorenz}, \citenamefont {Chen},
  \citenamefont {Guloy},\ and\ \citenamefont {Chu}}]{ThermopowerFeAS}%
  \BibitemOpen
  \bibfield  {author} {\bibinfo {author} {\bibfnamefont {B.}~\bibnamefont
  {Lv}}, \bibinfo {author} {\bibfnamefont {M.}~\bibnamefont {Gooch}}, \bibinfo
  {author} {\bibfnamefont {B.}~\bibnamefont {Lorenz}}, \bibinfo {author}
  {\bibfnamefont {F.}~\bibnamefont {Chen}}, \bibinfo {author} {\bibfnamefont
  {A.}~\bibnamefont {Guloy}}, \ and\ \bibinfo {author} {\bibfnamefont
  {C.}~\bibnamefont {Chu}},\ }\href@noop {} {\bibfield  {journal} {\bibinfo
  {journal} {New J. Phys.}\ }\textbf {\bibinfo {volume} {11}},\ \bibinfo
  {pages} {025013} (\bibinfo {year} {2009})}\BibitemShut {NoStop}%
\bibitem [{\citenamefont {L\"ohneysen}\ \emph {et~al.}(2007)\citenamefont
  {L\"ohneysen}, \citenamefont {Rosch}, \citenamefont {Vojta}, ,\ and\
  \citenamefont {W\''olfle}}]{HvLRMP2007}%
  \BibitemOpen
  \bibfield  {author} {\bibinfo {author} {\bibfnamefont {H.}~\bibnamefont
  {L\"ohneysen}}, \bibinfo {author} {\bibfnamefont {A.}~\bibnamefont {Rosch}},
  \bibinfo {author} {\bibfnamefont {M.}~\bibnamefont {Vojta}}, , \ and\
  \bibinfo {author} {\bibfnamefont {P.}~\bibnamefont {W\''olfle}},\ }\href@noop
  {} {\bibfield  {journal} {\bibinfo  {journal} {Rev. Mod. Phys.}\ }\textbf
  {\bibinfo {volume} {79}},\ \bibinfo {pages} {1015} (\bibinfo {year}
  {2007})}\BibitemShut {NoStop}%
\bibitem [{\citenamefont {Gegenwart}\ \emph {et~al.}(2007)\citenamefont
  {Gegenwart}, \citenamefont {Westerkamp}, \citenamefont {Krellner},
  \citenamefont {Tokiwa}, \citenamefont {Paschen}, \citenamefont {Geibel},
  \citenamefont {Steglich}, \citenamefont {Abrahams},\ and\ \citenamefont
  {Si}}]{PaschenYBRh}%
  \BibitemOpen
  \bibfield  {author} {\bibinfo {author} {\bibfnamefont {P.}~\bibnamefont
  {Gegenwart}}, \bibinfo {author} {\bibfnamefont {T.}~\bibnamefont
  {Westerkamp}}, \bibinfo {author} {\bibfnamefont {C.}~\bibnamefont
  {Krellner}}, \bibinfo {author} {\bibfnamefont {Y.}~\bibnamefont {Tokiwa}},
  \bibinfo {author} {\bibfnamefont {S.}~\bibnamefont {Paschen}}, \bibinfo
  {author} {\bibfnamefont {C.}~\bibnamefont {Geibel}}, \bibinfo {author}
  {\bibfnamefont {F.}~\bibnamefont {Steglich}}, \bibinfo {author}
  {\bibfnamefont {E.}~\bibnamefont {Abrahams}}, \ and\ \bibinfo {author}
  {\bibfnamefont {Q.}~\bibnamefont {Si}},\ }\href {\doibase
  10.1126/science.1136020} {\bibfield  {journal} {\bibinfo  {journal}
  {Science}\ }\textbf {\bibinfo {volume} {315}},\ \bibinfo {pages} {969}
  (\bibinfo {year} {2007})},\ \Eprint
  {http://arxiv.org/abs/https://www.science.org/doi/pdf/10.1126/science.1136020}
  {https://www.science.org/doi/pdf/10.1126/science.1136020} \BibitemShut
  {NoStop}%
\bibitem [{\citenamefont {Cao}\ \emph {et~al.}(2019)\citenamefont {Cao},
  \citenamefont {Chowdhury}, \citenamefont {Rodan-Legrain}, \citenamefont
  {Rubies-Bigorda}, \citenamefont {Watanabe}, \citenamefont {Taniguchi},
  \citenamefont {Senthil},\ and\ \citenamefont
  {Jarillo-Herrero}}]{Herrero2019}%
  \BibitemOpen
  \bibfield  {author} {\bibinfo {author} {\bibfnamefont {Y.}~\bibnamefont
  {Cao}}, \bibinfo {author} {\bibfnamefont {D.}~\bibnamefont {Chowdhury}},
  \bibinfo {author} {\bibfnamefont {D.}~\bibnamefont {Rodan-Legrain}}, \bibinfo
  {author} {\bibfnamefont {O.}~\bibnamefont {Rubies-Bigorda}}, \bibinfo
  {author} {\bibfnamefont {K.}~\bibnamefont {Watanabe}}, \bibinfo {author}
  {\bibfnamefont {T.}~\bibnamefont {Taniguchi}}, \bibinfo {author}
  {\bibfnamefont {T.}~\bibnamefont {Senthil}}, \ and\ \bibinfo {author}
  {\bibfnamefont {P.}~\bibnamefont {Jarillo-Herrero}},\ }\href@noop {}
  {\bibfield  {journal} {\bibinfo  {journal} {arXiv:1901.03710}\ } (\bibinfo
  {year} {2019})}\BibitemShut {NoStop}%
\bibitem [{\citenamefont {Lu}\ \emph {et~al.}(2019)\citenamefont {Lu},
  \citenamefont {Stepanov}, \citenamefont {Yang}, \citenamefont {Xie},
  \citenamefont {Aamir}, \citenamefont {Das}, \citenamefont {Urgell},
  \citenamefont {Watanabe}, \citenamefont {Zhang}, \citenamefont {Bachtold},
  \citenamefont {MacDonald},\ and\ \citenamefont {Efetov}}]{Efetov2019}%
  \BibitemOpen
  \bibfield  {author} {\bibinfo {author} {\bibfnamefont {X.}~\bibnamefont
  {Lu}}, \bibinfo {author} {\bibfnamefont {P.}~\bibnamefont {Stepanov}},
  \bibinfo {author} {\bibfnamefont {W.}~\bibnamefont {Yang}}, \bibinfo {author}
  {\bibfnamefont {M.}~\bibnamefont {Xie}}, \bibinfo {author} {\bibfnamefont
  {M.~A.}\ \bibnamefont {Aamir}}, \bibinfo {author} {\bibfnamefont
  {I.}~\bibnamefont {Das}}, \bibinfo {author} {\bibfnamefont {C.}~\bibnamefont
  {Urgell}}, \bibinfo {author} {\bibfnamefont {T.}~\bibnamefont {Watanabe},
  \bibfnamefont {Kenji~andTaniguchi}}, \bibinfo {author} {\bibfnamefont
  {G.}~\bibnamefont {Zhang}}, \bibinfo {author} {\bibfnamefont
  {A.}~\bibnamefont {Bachtold}}, \bibinfo {author} {\bibfnamefont {A.~H.}\
  \bibnamefont {MacDonald}}, \ and\ \bibinfo {author} {\bibfnamefont {D.~K.}\
  \bibnamefont {Efetov}},\ }\href@noop {} {\bibfield  {journal} {\bibinfo
  {journal} {arXiv:1903.06513}\ } (\bibinfo {year} {2019})}\BibitemShut
  {NoStop}%
\bibitem [{\citenamefont {Ghiotto}\ \emph {et~al.}(2021)\citenamefont
  {Ghiotto}, \citenamefont {Shih}, \citenamefont {Pereira}, \citenamefont
  {Rhodes}, \citenamefont {Kim}, \citenamefont {Zang}, \citenamefont {Millis},
  \citenamefont {Watanabe}, \citenamefont {Taniguchi}, \citenamefont {Hone},
  \citenamefont {Wang}, \citenamefont {Dean},\ and\ \citenamefont
  {Pasupathy}}]{Pasupathy2021}%
  \BibitemOpen
  \bibfield  {author} {\bibinfo {author} {\bibfnamefont {A.}~\bibnamefont
  {Ghiotto}}, \bibinfo {author} {\bibfnamefont {E.-M.}\ \bibnamefont {Shih}},
  \bibinfo {author} {\bibfnamefont {G.~S. S.~G.}\ \bibnamefont {Pereira}},
  \bibinfo {author} {\bibfnamefont {D.~A.}\ \bibnamefont {Rhodes}}, \bibinfo
  {author} {\bibfnamefont {B.}~\bibnamefont {Kim}}, \bibinfo {author}
  {\bibfnamefont {J.}~\bibnamefont {Zang}}, \bibinfo {author} {\bibfnamefont
  {A.~J.}\ \bibnamefont {Millis}}, \bibinfo {author} {\bibfnamefont
  {K.}~\bibnamefont {Watanabe}}, \bibinfo {author} {\bibfnamefont
  {T.}~\bibnamefont {Taniguchi}}, \bibinfo {author} {\bibfnamefont {J.~C.}\
  \bibnamefont {Hone}}, \bibinfo {author} {\bibfnamefont {L.}~\bibnamefont
  {Wang}}, \bibinfo {author} {\bibfnamefont {C.~R.}\ \bibnamefont {Dean}}, \
  and\ \bibinfo {author} {\bibfnamefont {A.~N.}\ \bibnamefont {Pasupathy}},\
  }\href@noop {} {\  (\bibinfo {year} {2021})},\ \Eprint
  {http://arxiv.org/abs/Preprint available at http://arxiv.org/abs/2103.09796}
  {arXiv:Preprint available at http://arxiv.org/abs/2103.09796} \BibitemShut
  {NoStop}%
\bibitem [{\citenamefont {Hayes}\ \emph {et~al.}(2014)\citenamefont {Hayes},
  \citenamefont {Breznay}, \citenamefont {Helm}, \citenamefont {Moll},
  \citenamefont {Wartenbe}, \citenamefont {McDonald}, \citenamefont
  {Shekhter},\ and\ \citenamefont {Analytis}}]{Analytis1}%
  \BibitemOpen
  \bibfield  {author} {\bibinfo {author} {\bibfnamefont {I.}~\bibnamefont
  {Hayes}}, \bibinfo {author} {\bibfnamefont {N.}~\bibnamefont {Breznay}},
  \bibinfo {author} {\bibfnamefont {T.}~\bibnamefont {Helm}}, \bibinfo {author}
  {\bibfnamefont {P.}~\bibnamefont {Moll}}, \bibinfo {author} {\bibfnamefont
  {B.}~\bibnamefont {Wartenbe}}, \bibinfo {author} {\bibfnamefont
  {R.}~\bibnamefont {McDonald}}, \bibinfo {author} {\bibfnamefont
  {A.}~\bibnamefont {Shekhter}}, \ and\ \bibinfo {author} {\bibfnamefont
  {J.~G.}\ \bibnamefont {Analytis}},\ }\href@noop {} {\bibfield  {journal}
  {\bibinfo  {journal} {arXiv:1412.6484 [cond-mat.str-el]}\ } (\bibinfo {year}
  {2014})}\BibitemShut {NoStop}%
\bibitem [{\citenamefont {Giraldo-Gallo}\ \emph {et~al.}(2018)\citenamefont
  {Giraldo-Gallo}, \citenamefont {Galvis}, \citenamefont {Stegen},
  \citenamefont {Modic}, \citenamefont {Balakirev}, \citenamefont {Betts},
  \citenamefont {Lian}, \citenamefont {Moir}, \citenamefont {Riggs},
  \citenamefont {Wu}, \citenamefont {Bollinger}, \citenamefont {He},
  \citenamefont {Bo{\v z}ovi{\'c}}, \citenamefont {Ramshaw}, \citenamefont
  {McDonald}, \citenamefont {Boebinger},\ and\ \citenamefont
  {Shekhter}}]{Shekhter2018}%
  \BibitemOpen
  \bibfield  {author} {\bibinfo {author} {\bibfnamefont {P.}~\bibnamefont
  {Giraldo-Gallo}}, \bibinfo {author} {\bibfnamefont {J.~A.}\ \bibnamefont
  {Galvis}}, \bibinfo {author} {\bibfnamefont {Z.}~\bibnamefont {Stegen}},
  \bibinfo {author} {\bibfnamefont {K.~A.}\ \bibnamefont {Modic}}, \bibinfo
  {author} {\bibfnamefont {F.~F.}\ \bibnamefont {Balakirev}}, \bibinfo {author}
  {\bibfnamefont {J.~B.}\ \bibnamefont {Betts}}, \bibinfo {author}
  {\bibfnamefont {X.}~\bibnamefont {Lian}}, \bibinfo {author} {\bibfnamefont
  {C.}~\bibnamefont {Moir}}, \bibinfo {author} {\bibfnamefont {S.~C.}\
  \bibnamefont {Riggs}}, \bibinfo {author} {\bibfnamefont {J.}~\bibnamefont
  {Wu}}, \bibinfo {author} {\bibfnamefont {A.~T.}\ \bibnamefont {Bollinger}},
  \bibinfo {author} {\bibfnamefont {X.}~\bibnamefont {He}}, \bibinfo {author}
  {\bibfnamefont {I.}~\bibnamefont {Bo{\v z}ovi{\'c}}}, \bibinfo {author}
  {\bibfnamefont {B.~J.}\ \bibnamefont {Ramshaw}}, \bibinfo {author}
  {\bibfnamefont {R.~D.}\ \bibnamefont {McDonald}}, \bibinfo {author}
  {\bibfnamefont {G.~S.}\ \bibnamefont {Boebinger}}, \ and\ \bibinfo {author}
  {\bibfnamefont {A.}~\bibnamefont {Shekhter}},\ }\href {\doibase
  10.1126/science.aan3178} {\bibfield  {journal} {\bibinfo  {journal}
  {Science}\ }\textbf {\bibinfo {volume} {361}},\ \bibinfo {pages} {479}
  (\bibinfo {year} {2018})},\ \Eprint
  {http://arxiv.org/abs/https://science.sciencemag.org/content/361/6401/479.full.pdf}
  {https://science.sciencemag.org/content/361/6401/479.full.pdf} \BibitemShut
  {NoStop}%
\bibitem [{\citenamefont {Jaoui}\ \emph {et~al.}(2021)\citenamefont {Jaoui},
  \citenamefont {Das}, \citenamefont {Di~Battista}, \citenamefont
  {D{\'\i}ez-M{\'e}rida}, \citenamefont {Lu}, \citenamefont {Watanabe},
  \citenamefont {Taniguchi}, \citenamefont {Ishizuka}, \citenamefont
  {Levitov},\ and\ \citenamefont {Efetov}}]{Efetov2021}%
  \BibitemOpen
  \bibfield  {author} {\bibinfo {author} {\bibfnamefont {A.}~\bibnamefont
  {Jaoui}}, \bibinfo {author} {\bibfnamefont {I.}~\bibnamefont {Das}}, \bibinfo
  {author} {\bibfnamefont {G.}~\bibnamefont {Di~Battista}}, \bibinfo {author}
  {\bibfnamefont {J.}~\bibnamefont {D{\'\i}ez-M{\'e}rida}}, \bibinfo {author}
  {\bibfnamefont {X.}~\bibnamefont {Lu}}, \bibinfo {author} {\bibfnamefont
  {K.}~\bibnamefont {Watanabe}}, \bibinfo {author} {\bibfnamefont
  {T.}~\bibnamefont {Taniguchi}}, \bibinfo {author} {\bibfnamefont
  {H.}~\bibnamefont {Ishizuka}}, \bibinfo {author} {\bibfnamefont
  {L.}~\bibnamefont {Levitov}}, \ and\ \bibinfo {author} {\bibfnamefont
  {D.~K.}\ \bibnamefont {Efetov}},\ }\href@noop {} {\  (\bibinfo {year}
  {2021})},\ \Eprint {http://arxiv.org/abs/Preprint available at
  http://arxiv.org/abs/2108.07753} {arXiv:Preprint available at
  http://arxiv.org/abs/2108.07753} \BibitemShut {NoStop}%
\bibitem [{\citenamefont {Hayes}\ \emph {et~al.}(2018)\citenamefont {Hayes},
  \citenamefont {Hao}, \citenamefont {Maksimovic}, \citenamefont {Lewin},
  \citenamefont {Chan}, \citenamefont {McDonald}, \citenamefont {Ramshaw},
  \citenamefont {Moore},\ and\ \citenamefont {Analytis}}]{Hayes2018}%
  \BibitemOpen
  \bibfield  {author} {\bibinfo {author} {\bibfnamefont {I.~M.}\ \bibnamefont
  {Hayes}}, \bibinfo {author} {\bibfnamefont {Z.}~\bibnamefont {Hao}}, \bibinfo
  {author} {\bibfnamefont {N.}~\bibnamefont {Maksimovic}}, \bibinfo {author}
  {\bibfnamefont {S.~K.}\ \bibnamefont {Lewin}}, \bibinfo {author}
  {\bibfnamefont {M.~K.}\ \bibnamefont {Chan}}, \bibinfo {author}
  {\bibfnamefont {R.~D.}\ \bibnamefont {McDonald}}, \bibinfo {author}
  {\bibfnamefont {B.~J.}\ \bibnamefont {Ramshaw}}, \bibinfo {author}
  {\bibfnamefont {J.~E.}\ \bibnamefont {Moore}}, \ and\ \bibinfo {author}
  {\bibfnamefont {J.~G.}\ \bibnamefont {Analytis}},\ }\href@noop {} {\bibfield
  {journal} {\bibinfo  {journal} {Phys. Rev. Lett.}\ }\textbf {\bibinfo
  {volume} {121}},\ \bibinfo {pages} {197002} (\bibinfo {year}
  {2018})}\BibitemShut {NoStop}%
\bibitem [{Note1()}]{Note1}%
  \BibitemOpen
  \bibinfo {note} {G. Boebinger and A. Shekhter - private communication
  2020}\BibitemShut {NoStop}%
\bibitem [{\citenamefont {Aji}\ and\ \citenamefont {Varma}(2007)}]{Aji-V-qcf1}%
  \BibitemOpen
  \bibfield  {author} {\bibinfo {author} {\bibfnamefont {V.}~\bibnamefont
  {Aji}}\ and\ \bibinfo {author} {\bibfnamefont {C.~M.}\ \bibnamefont
  {Varma}},\ }\href {\doibase 10.1103/PhysRevLett.99.067003} {\bibfield
  {journal} {\bibinfo  {journal} {Phys. Rev. Lett.}\ }\textbf {\bibinfo
  {volume} {99}},\ \bibinfo {pages} {067003} (\bibinfo {year}
  {2007})}\BibitemShut {NoStop}%
\bibitem [{\citenamefont {Aji}\ and\ \citenamefont {Varma}(2010)}]{Aji-V-qcf3}%
  \BibitemOpen
  \bibfield  {author} {\bibinfo {author} {\bibfnamefont {V.}~\bibnamefont
  {Aji}}\ and\ \bibinfo {author} {\bibfnamefont {C.~M.}\ \bibnamefont
  {Varma}},\ }\href {\doibase 10.1103/PhysRevB.82.174501} {\bibfield  {journal}
  {\bibinfo  {journal} {Phys. Rev. B}\ }\textbf {\bibinfo {volume} {82}},\
  \bibinfo {pages} {174501} (\bibinfo {year} {2010})}\BibitemShut {NoStop}%
\bibitem [{\citenamefont {Varma}(2012)}]{CMV-IOP-REV}%
  \BibitemOpen
  \bibfield  {author} {\bibinfo {author} {\bibfnamefont {C.~M.}\ \bibnamefont
  {Varma}},\ }\href@noop {} {\bibfield  {journal} {\bibinfo  {journal} {Rep.
  Prog. Phys.}\ }\textbf {\bibinfo {volume} {75}},\ \bibinfo {pages} {052501}
  (\bibinfo {year} {2012})}\BibitemShut {NoStop}%
\bibitem [{\citenamefont {Else}\ and\ \citenamefont
  {Senthil}(2021)}]{ElseSenthil2021}%
  \BibitemOpen
  \bibfield  {author} {\bibinfo {author} {\bibfnamefont {D.~V.}\ \bibnamefont
  {Else}}\ and\ \bibinfo {author} {\bibfnamefont {T.}~\bibnamefont {Senthil}},\
  }\href@noop {} {\bibfield  {journal} {\bibinfo  {journal} {Phys. Rev. Lett.}\
  }\textbf {\bibinfo {volume} {127}},\ \bibinfo {pages} {086601} (\bibinfo
  {year} {2021})}\BibitemShut {NoStop}%
\bibitem [{\citenamefont {Kaminski}\ \emph {et~al.}(2002)\citenamefont
  {Kaminski}, \citenamefont {Rosenkranz}, \citenamefont {Fretwell},
  \citenamefont {Campuzano}, \citenamefont {Li}, \citenamefont {Raffy},
  \citenamefont {Cullen}, \citenamefont {You}, \citenamefont {Olson},
  \citenamefont {Varma},\ and\ \citenamefont {H\"{o}chst}}]{Kaminski-diARPES}%
  \BibitemOpen
  \bibfield  {author} {\bibinfo {author} {\bibfnamefont {A.}~\bibnamefont
  {Kaminski}}, \bibinfo {author} {\bibfnamefont {S.}~\bibnamefont
  {Rosenkranz}}, \bibinfo {author} {\bibfnamefont {H.~M.}\ \bibnamefont
  {Fretwell}}, \bibinfo {author} {\bibfnamefont {J.~C.}\ \bibnamefont
  {Campuzano}}, \bibinfo {author} {\bibfnamefont {Z.}~\bibnamefont {Li}},
  \bibinfo {author} {\bibfnamefont {H.}~\bibnamefont {Raffy}}, \bibinfo
  {author} {\bibfnamefont {W.~G.}\ \bibnamefont {Cullen}}, \bibinfo {author}
  {\bibfnamefont {H.}~\bibnamefont {You}}, \bibinfo {author} {\bibfnamefont
  {C.~G.}\ \bibnamefont {Olson}}, \bibinfo {author} {\bibfnamefont {C.~M.}\
  \bibnamefont {Varma}}, \ and\ \bibinfo {author} {\bibfnamefont
  {H.}~\bibnamefont {H\"{o}chst}},\ }\href {\doibase 10.1038/416610a}
  {\bibfield  {journal} {\bibinfo  {journal} {Nature}\ }\textbf {\bibinfo
  {volume} {416}},\ \bibinfo {pages} {610} (\bibinfo {year}
  {2002})}\BibitemShut {NoStop}%
\bibitem [{\citenamefont {Bourges}\ and\ \citenamefont
  {Sidis}(2011)}]{Bourges-rev}%
  \BibitemOpen
  \bibfield  {author} {\bibinfo {author} {\bibfnamefont {P.}~\bibnamefont
  {Bourges}}\ and\ \bibinfo {author} {\bibfnamefont {Y.}~\bibnamefont
  {Sidis}},\ }\href {\doibase http://dx.doi.org/10.1016/j.crhy.2011.04.006}
  {\bibfield  {journal} {\bibinfo  {journal} {Comptes Rendus Physique}\
  }\textbf {\bibinfo {volume} {12}},\ \bibinfo {pages} {461 } (\bibinfo {year}
  {2011})}\BibitemShut {NoStop}%
\bibitem [{\citenamefont {Zhao}\ \emph {et~al.}(2016)\citenamefont {Zhao},
  \citenamefont {Belvin}, \citenamefont {Liang}, \citenamefont {Bonn},
  \citenamefont {Hardy}, \citenamefont {Armitage},\ and\ \citenamefont
  {Hsieh}}]{Hsieh2017}%
  \BibitemOpen
  \bibfield  {author} {\bibinfo {author} {\bibfnamefont {L.}~\bibnamefont
  {Zhao}}, \bibinfo {author} {\bibfnamefont {C.~A.}\ \bibnamefont {Belvin}},
  \bibinfo {author} {\bibfnamefont {R.}~\bibnamefont {Liang}}, \bibinfo
  {author} {\bibfnamefont {D.~A.}\ \bibnamefont {Bonn}}, \bibinfo {author}
  {\bibfnamefont {W.~N.}\ \bibnamefont {Hardy}}, \bibinfo {author}
  {\bibfnamefont {N.~P.}\ \bibnamefont {Armitage}}, \ and\ \bibinfo {author}
  {\bibfnamefont {D.}~\bibnamefont {Hsieh}},\ }\href@noop {} {\bibfield
  {journal} {\bibinfo  {journal} {Nature Physics}\ }\textbf {\bibinfo {volume}
  {13}},\ \bibinfo {pages} {250 EP} (\bibinfo {year} {2016})}\BibitemShut
  {NoStop}%
\bibitem [{\citenamefont {Lubashevsky}\ \emph {et~al.}(2014)\citenamefont
  {Lubashevsky}, \citenamefont {Pan}, \citenamefont {Kirzhner}, \citenamefont
  {Koren},\ and\ \citenamefont {Armitage}}]{Armitage-Biref}%
  \BibitemOpen
  \bibfield  {author} {\bibinfo {author} {\bibfnamefont {Y.}~\bibnamefont
  {Lubashevsky}}, \bibinfo {author} {\bibfnamefont {L.}~\bibnamefont {Pan}},
  \bibinfo {author} {\bibfnamefont {T.}~\bibnamefont {Kirzhner}}, \bibinfo
  {author} {\bibfnamefont {G.}~\bibnamefont {Koren}}, \ and\ \bibinfo {author}
  {\bibfnamefont {N.~P.}\ \bibnamefont {Armitage}},\ }\href {\doibase
  10.1103/PhysRevLett.112.147001} {\bibfield  {journal} {\bibinfo  {journal}
  {Phys. Rev. Lett.}\ }\textbf {\bibinfo {volume} {112}},\ \bibinfo {pages}
  {147001} (\bibinfo {year} {2014})}\BibitemShut {NoStop}%
\bibitem [{\citenamefont {Zhang}\ \emph {et~al.}(2018)\citenamefont {Zhang},
  \citenamefont {Ding}, \citenamefont {Tan}, \citenamefont {Huang},
  \citenamefont {Bernal}, \citenamefont {Ho}, \citenamefont {Morris},
  \citenamefont {Hillier}, \citenamefont {Biswas}, \citenamefont {Cottrell},
  \citenamefont {Xiang}, \citenamefont {Yao}, \citenamefont {MacLaughlin},\
  and\ \citenamefont {Shu}}]{Shu2018}%
  \BibitemOpen
  \bibfield  {author} {\bibinfo {author} {\bibfnamefont {J.}~\bibnamefont
  {Zhang}}, \bibinfo {author} {\bibfnamefont {Z.}~\bibnamefont {Ding}},
  \bibinfo {author} {\bibfnamefont {C.}~\bibnamefont {Tan}}, \bibinfo {author}
  {\bibfnamefont {K.}~\bibnamefont {Huang}}, \bibinfo {author} {\bibfnamefont
  {O.~O.}\ \bibnamefont {Bernal}}, \bibinfo {author} {\bibfnamefont {P.-C.}\
  \bibnamefont {Ho}}, \bibinfo {author} {\bibfnamefont {G.~D.}\ \bibnamefont
  {Morris}}, \bibinfo {author} {\bibfnamefont {A.~D.}\ \bibnamefont {Hillier}},
  \bibinfo {author} {\bibfnamefont {P.~K.}\ \bibnamefont {Biswas}}, \bibinfo
  {author} {\bibfnamefont {S.~P.}\ \bibnamefont {Cottrell}}, \bibinfo {author}
  {\bibfnamefont {H.}~\bibnamefont {Xiang}}, \bibinfo {author} {\bibfnamefont
  {X.}~\bibnamefont {Yao}}, \bibinfo {author} {\bibfnamefont {D.~E.}\
  \bibnamefont {MacLaughlin}}, \ and\ \bibinfo {author} {\bibfnamefont
  {L.}~\bibnamefont {Shu}},\ }\href {\doibase 10.1126/sciadv.aao5235}
  {\bibfield  {journal} {\bibinfo  {journal} {Science Advances}\ }\textbf
  {\bibinfo {volume} {4}} (\bibinfo {year} {2018}),\ 10.1126/sciadv.aao5235},\
  \Eprint
  {http://arxiv.org/abs/https://advances.sciencemag.org/content/4/1/eaao5235.full.pdf}
  {https://advances.sciencemag.org/content/4/1/eaao5235.full.pdf} \BibitemShut
  {NoStop}%
\bibitem [{\citenamefont {Zhu}\ \emph {et~al.}(2015)\citenamefont {Zhu},
  \citenamefont {Chen},\ and\ \citenamefont {Varma}}]{ZhuChenCMV2015}%
  \BibitemOpen
  \bibfield  {author} {\bibinfo {author} {\bibfnamefont {L.}~\bibnamefont
  {Zhu}}, \bibinfo {author} {\bibfnamefont {Y.}~\bibnamefont {Chen}}, \ and\
  \bibinfo {author} {\bibfnamefont {C.~M.}\ \bibnamefont {Varma}},\ }\href
  {\doibase 10.1103/PhysRevB.91.205129} {\bibfield  {journal} {\bibinfo
  {journal} {Phys. Rev. B}\ }\textbf {\bibinfo {volume} {91}},\ \bibinfo
  {pages} {205129} (\bibinfo {year} {2015})}\BibitemShut {NoStop}%
\bibitem [{\citenamefont {Zhu}\ \emph {et~al.}(2016)\citenamefont {Zhu},
  \citenamefont {Hou},\ and\ \citenamefont {Varma}}]{ZhuHouV2016}%
  \BibitemOpen
  \bibfield  {author} {\bibinfo {author} {\bibfnamefont {L.}~\bibnamefont
  {Zhu}}, \bibinfo {author} {\bibfnamefont {C.}~\bibnamefont {Hou}}, \ and\
  \bibinfo {author} {\bibfnamefont {C.~M.}\ \bibnamefont {Varma}},\ }\href
  {\doibase 10.1103/PhysRevB.94.235156} {\bibfield  {journal} {\bibinfo
  {journal} {Phys. Rev. B}\ }\textbf {\bibinfo {volume} {94}},\ \bibinfo
  {pages} {235156} (\bibinfo {year} {2016})}\BibitemShut {NoStop}%
\bibitem [{\citenamefont {Hou}\ and\ \citenamefont {Varma}(2016)}]{Hou-CMV-RG}%
  \BibitemOpen
  \bibfield  {author} {\bibinfo {author} {\bibfnamefont {C.}~\bibnamefont
  {Hou}}\ and\ \bibinfo {author} {\bibfnamefont {C.~M.}\ \bibnamefont
  {Varma}},\ }\href {\doibase 10.1103/PhysRevB.94.201101} {\bibfield  {journal}
  {\bibinfo  {journal} {Phys. Rev. B}\ }\textbf {\bibinfo {volume} {94}},\
  \bibinfo {pages} {201101} (\bibinfo {year} {2016})}\BibitemShut {NoStop}%
\bibitem [{\citenamefont {Aji}\ \emph {et~al.}(2010)\citenamefont {Aji},
  \citenamefont {Shekhter},\ and\ \citenamefont {Varma}}]{ASV2010}%
  \BibitemOpen
  \bibfield  {author} {\bibinfo {author} {\bibfnamefont {V.}~\bibnamefont
  {Aji}}, \bibinfo {author} {\bibfnamefont {A.}~\bibnamefont {Shekhter}}, \
  and\ \bibinfo {author} {\bibfnamefont {C.~M.}\ \bibnamefont {Varma}},\ }\href
  {\doibase 10.1103/PhysRevB.81.064515} {\bibfield  {journal} {\bibinfo
  {journal} {Phys. Rev. B}\ }\textbf {\bibinfo {volume} {81}},\ \bibinfo
  {pages} {064515} (\bibinfo {year} {2010})}\BibitemShut {NoStop}%
\bibitem [{\citenamefont {Varma}\ \emph {et~al.}(1989)\citenamefont {Varma},
  \citenamefont {Littlewood}, \citenamefont {Schmitt-Rink}, \citenamefont
  {Abrahams},\ and\ \citenamefont {Ruckenstein}}]{CMV-MFL}%
  \BibitemOpen
  \bibfield  {author} {\bibinfo {author} {\bibfnamefont {C.~M.}\ \bibnamefont
  {Varma}}, \bibinfo {author} {\bibfnamefont {P.~B.}\ \bibnamefont
  {Littlewood}}, \bibinfo {author} {\bibfnamefont {S.}~\bibnamefont
  {Schmitt-Rink}}, \bibinfo {author} {\bibfnamefont {E.}~\bibnamefont
  {Abrahams}}, \ and\ \bibinfo {author} {\bibfnamefont {A.~E.}\ \bibnamefont
  {Ruckenstein}},\ }\href {\doibase 10.1103/PhysRevLett.63.1996} {\bibfield
  {journal} {\bibinfo  {journal} {Phys. Rev. Lett.}\ }\textbf {\bibinfo
  {volume} {63}},\ \bibinfo {pages} {1996} (\bibinfo {year}
  {1989})}\BibitemShut {NoStop}%
\bibitem [{\citenamefont {Chowdhury}\ \emph {et~al.}(2021)\citenamefont
  {Chowdhury}, \citenamefont {Georges}, \citenamefont {Parcollet},\ and\
  \citenamefont {Sachdev}}]{SYK2021}%
  \BibitemOpen
  \bibfield  {author} {\bibinfo {author} {\bibfnamefont {D.}~\bibnamefont
  {Chowdhury}}, \bibinfo {author} {\bibfnamefont {A.}~\bibnamefont {Georges}},
  \bibinfo {author} {\bibfnamefont {O.}~\bibnamefont {Parcollet}}, \ and\
  \bibinfo {author} {\bibfnamefont {S.}~\bibnamefont {Sachdev}},\ }\href@noop
  {} {\  (\bibinfo {year} {2021})},\ \Eprint {http://arxiv.org/abs/Preprint
  available at http://arxiv.org/abs/2109.05037} {arXiv:Preprint available at
  http://arxiv.org/abs/2109.05037} \BibitemShut {NoStop}%
\bibitem [{\citenamefont {Faulkner}\ \emph {et~al.}(2011)\citenamefont
  {Faulkner}, \citenamefont {Liu}, \citenamefont {McGreevy},\ and\
  \citenamefont {Vegh}}]{HongLiu2011}%
  \BibitemOpen
  \bibfield  {author} {\bibinfo {author} {\bibfnamefont {T.}~\bibnamefont
  {Faulkner}}, \bibinfo {author} {\bibfnamefont {H.}~\bibnamefont {Liu}},
  \bibinfo {author} {\bibfnamefont {J.}~\bibnamefont {McGreevy}}, \ and\
  \bibinfo {author} {\bibfnamefont {D.}~\bibnamefont {Vegh}},\ }\href {\doibase
  10.1103/PhysRevD.83.125002} {\bibfield  {journal} {\bibinfo  {journal} {Phys.
  Rev. D}\ }\textbf {\bibinfo {volume} {83}},\ \bibinfo {pages} {125002}
  (\bibinfo {year} {2011})}\BibitemShut {NoStop}%
\bibitem [{\citenamefont {Dumitrescu}\ \emph {et~al.}(2021)\citenamefont
  {Dumitrescu}, \citenamefont {Wentzell}, \citenamefont {Georges},\ and\
  \citenamefont {Parcollet}}]{Dumeitrescu2021}%
  \BibitemOpen
  \bibfield  {author} {\bibinfo {author} {\bibfnamefont {P.~T.}\ \bibnamefont
  {Dumitrescu}}, \bibinfo {author} {\bibfnamefont {N.}~\bibnamefont
  {Wentzell}}, \bibinfo {author} {\bibfnamefont {A.}~\bibnamefont {Georges}}, \
  and\ \bibinfo {author} {\bibfnamefont {O.}~\bibnamefont {Parcollet}},\
  }\href@noop {} {\  (\bibinfo {year} {2021})},\ \Eprint
  {http://arxiv.org/abs/Preprint available at http://arxiv.org/abs/2103.08607}
  {arXiv:Preprint available at http://arxiv.org/abs/2103.08607} \BibitemShut
  {NoStop}%
\bibitem [{\citenamefont {Ruckenstein}\ and\ \citenamefont
  {Varma}(1991)}]{Ruckenstein1991}%
  \BibitemOpen
  \bibfield  {author} {\bibinfo {author} {\bibfnamefont {A.~E.}\ \bibnamefont
  {Ruckenstein}}\ and\ \bibinfo {author} {\bibfnamefont {C.~M.}\ \bibnamefont
  {Varma}},\ }\href {\doibase 10.1016/0921-4534(91)91962-4} {\bibfield
  {journal} {\bibinfo  {journal} {Physica C Superconductivity}\ }\textbf
  {\bibinfo {volume} {185}},\ \bibinfo {pages} {134} (\bibinfo {year}
  {1991})}\BibitemShut {NoStop}%
\bibitem [{\citenamefont {Mousatov}\ \emph {et~al.}(2020)\citenamefont
  {Mousatov}, \citenamefont {Berg},\ and\ \citenamefont
  {Hartnoll}}]{BergHartnoll2020}%
  \BibitemOpen
  \bibfield  {author} {\bibinfo {author} {\bibfnamefont {C.~H.}\ \bibnamefont
  {Mousatov}}, \bibinfo {author} {\bibfnamefont {E.}~\bibnamefont {Berg}}, \
  and\ \bibinfo {author} {\bibfnamefont {S.~A.}\ \bibnamefont {Hartnoll}},\
  }\href {\doibase 10.1073/pnas.1915224117} {\bibfield  {journal} {\bibinfo
  {journal} {Proceedings of the National Academy of Sciences}\ }\textbf
  {\bibinfo {volume} {117}},\ \bibinfo {pages} {2852} (\bibinfo {year}
  {2020})},\ \Eprint
  {http://arxiv.org/abs/https://www.pnas.org/content/117/6/2852.full.pdf}
  {https://www.pnas.org/content/117/6/2852.full.pdf} \BibitemShut {NoStop}%
\bibitem [{\citenamefont {Sun}\ and\ \citenamefont
  {Fradkin}(2008)}]{SunFradkin2008}%
  \BibitemOpen
  \bibfield  {author} {\bibinfo {author} {\bibfnamefont {K.}~\bibnamefont
  {Sun}}\ and\ \bibinfo {author} {\bibfnamefont {E.}~\bibnamefont {Fradkin}},\
  }\href {\doibase 10.1103/PhysRevB.78.245122} {\bibfield  {journal} {\bibinfo
  {journal} {Phys. Rev. B}\ }\textbf {\bibinfo {volume} {78}},\ \bibinfo
  {pages} {245122} (\bibinfo {year} {2008})}\BibitemShut {NoStop}%
\bibitem [{\citenamefont {He}\ and\ \citenamefont
  {Varma}(2012)}]{He-V-neutrons}%
  \BibitemOpen
  \bibfield  {author} {\bibinfo {author} {\bibfnamefont {Y.}~\bibnamefont
  {He}}\ and\ \bibinfo {author} {\bibfnamefont {C.~M.}\ \bibnamefont {Varma}},\
  }\href@noop {} {\bibfield  {journal} {\bibinfo  {journal} {Phys. Rev. B}\
  }\textbf {\bibinfo {volume} {86}},\ \bibinfo {pages} {035124} (\bibinfo
  {year} {2012})}\BibitemShut {NoStop}%
\bibitem [{\citenamefont {He}\ \emph {et~al.}(2012)\citenamefont {He},
  \citenamefont {Moore},\ and\ \citenamefont {Varma}}]{HeMooreV2012}%
  \BibitemOpen
  \bibfield  {author} {\bibinfo {author} {\bibfnamefont {Y.}~\bibnamefont
  {He}}, \bibinfo {author} {\bibfnamefont {J.}~\bibnamefont {Moore}}, \ and\
  \bibinfo {author} {\bibfnamefont {C.~M.}\ \bibnamefont {Varma}},\ }\href
  {\doibase 10.1103/PhysRevB.85.155106} {\bibfield  {journal} {\bibinfo
  {journal} {Phys. Rev. B}\ }\textbf {\bibinfo {volume} {85}},\ \bibinfo
  {pages} {155106} (\bibinfo {year} {2012})}\BibitemShut {NoStop}%
\bibitem [{\citenamefont {Pelzer}(1991)}]{Pelzer1991}%
  \BibitemOpen
  \bibfield  {author} {\bibinfo {author} {\bibfnamefont {F.}~\bibnamefont
  {Pelzer}},\ }\href {\doibase 10.1103/PhysRevB.44.293} {\bibfield  {journal}
  {\bibinfo  {journal} {Phys. Rev. B}\ }\textbf {\bibinfo {volume} {44}},\
  \bibinfo {pages} {293} (\bibinfo {year} {1991})}\BibitemShut {NoStop}%
\bibitem [{\citenamefont {Varma}\ \emph {et~al.}(2002)\citenamefont {Varma},
  \citenamefont {Nussinov},\ and\ \citenamefont {van Saarloos}}]{CMV_Lorentz}%
  \BibitemOpen
  \bibfield  {author} {\bibinfo {author} {\bibfnamefont {C.}~\bibnamefont
  {Varma}}, \bibinfo {author} {\bibfnamefont {Z.}~\bibnamefont {Nussinov}}, \
  and\ \bibinfo {author} {\bibfnamefont {W.}~\bibnamefont {van Saarloos}},\
  }\href@noop {} {\bibfield  {journal} {\bibinfo  {journal} {Phys. Reports}\
  }\textbf {\bibinfo {volume} {361}},\ \bibinfo {pages} {267} (\bibinfo {year}
  {2002})}\BibitemShut {NoStop}%
\bibitem [{Note2()}]{Note2}%
  \BibitemOpen
  \bibinfo {note} {H. Maebashi, C.M. Varma - unpublished}\BibitemShut {NoStop}%
\bibitem [{Note3()}]{Note3}%
  \BibitemOpen
  \bibinfo {note} {A. Shekhter - private communication (Dec. 2021)}\BibitemShut
  {NoStop}%
\bibitem [{\citenamefont {Abrahams}\ and\ \citenamefont
  {Varma}(2003)}]{Abrahams-VarmaPRB}%
  \BibitemOpen
  \bibfield  {author} {\bibinfo {author} {\bibfnamefont {E.}~\bibnamefont
  {Abrahams}}\ and\ \bibinfo {author} {\bibfnamefont {C.~M.}\ \bibnamefont
  {Varma}},\ }\href {\doibase 10.1103/PhysRevB.68.094502} {\bibfield  {journal}
  {\bibinfo  {journal} {Phys. Rev. B}\ }\textbf {\bibinfo {volume} {68}},\
  \bibinfo {pages} {094502} (\bibinfo {year} {2003})}\BibitemShut {NoStop}%
\bibitem [{\citenamefont {Michon}\ \emph {et~al.}(2019)\citenamefont {Michon},
  \citenamefont {Girod}, \citenamefont {Badoux}, \citenamefont {Ka{\v c}mar{\v
  c}{\'\i}k}, \citenamefont {Ma}, \citenamefont {Dragomir}, \citenamefont
  {Dabkowska}, \citenamefont {Gaulin}, \citenamefont {Zhou}, \citenamefont
  {Pyon}, \citenamefont {Takayama}, \citenamefont {Takagi}, \citenamefont
  {Verret}, \citenamefont {Doiron-Leyraud}, \citenamefont {Marcenat},
  \citenamefont {Taillefer},\ and\ \citenamefont {Klein}}]{Tailleferspht}%
  \BibitemOpen
  \bibfield  {author} {\bibinfo {author} {\bibfnamefont {B.}~\bibnamefont
  {Michon}}, \bibinfo {author} {\bibfnamefont {C.}~\bibnamefont {Girod}},
  \bibinfo {author} {\bibfnamefont {S.}~\bibnamefont {Badoux}}, \bibinfo
  {author} {\bibfnamefont {J.}~\bibnamefont {Ka{\v c}mar{\v c}{\'\i}k}},
  \bibinfo {author} {\bibfnamefont {Q.}~\bibnamefont {Ma}}, \bibinfo {author}
  {\bibfnamefont {M.}~\bibnamefont {Dragomir}}, \bibinfo {author}
  {\bibfnamefont {H.~A.}\ \bibnamefont {Dabkowska}}, \bibinfo {author}
  {\bibfnamefont {B.~D.}\ \bibnamefont {Gaulin}}, \bibinfo {author}
  {\bibfnamefont {J.-S.}\ \bibnamefont {Zhou}}, \bibinfo {author}
  {\bibfnamefont {S.}~\bibnamefont {Pyon}}, \bibinfo {author} {\bibfnamefont
  {T.}~\bibnamefont {Takayama}}, \bibinfo {author} {\bibfnamefont
  {H.}~\bibnamefont {Takagi}}, \bibinfo {author} {\bibfnamefont
  {S.}~\bibnamefont {Verret}}, \bibinfo {author} {\bibfnamefont
  {N.}~\bibnamefont {Doiron-Leyraud}}, \bibinfo {author} {\bibfnamefont
  {C.}~\bibnamefont {Marcenat}}, \bibinfo {author} {\bibfnamefont
  {L.}~\bibnamefont {Taillefer}}, \ and\ \bibinfo {author} {\bibfnamefont
  {T.}~\bibnamefont {Klein}},\ }\href@noop {} {\bibfield  {journal} {\bibinfo
  {journal} {Nature}\ }\textbf {\bibinfo {volume} {567}},\ \bibinfo {pages}
  {218} (\bibinfo {year} {2019})}\BibitemShut {NoStop}%
\bibitem [{\citenamefont {McMillan}(1975)}]{McMillanCDW}%
  \BibitemOpen
  \bibfield  {author} {\bibinfo {author} {\bibfnamefont {W.}~\bibnamefont
  {McMillan}},\ }\href@noop {} {\bibfield  {journal} {\bibinfo  {journal}
  {Phys. Rev. B}\ }\textbf {\bibinfo {volume} {12}},\ \bibinfo {pages} {1187}
  (\bibinfo {year} {1975})}\BibitemShut {NoStop}%
\bibitem [{\citenamefont {Inosov}\ \emph {et~al.}(2010)\citenamefont {Inosov},
  \citenamefont {Park}, \citenamefont {Bourges}, \citenamefont {Sun},
  \citenamefont {Sidis}, \citenamefont {Schneidewind}, \citenamefont {Hradil},
  \citenamefont {Haug}, \citenamefont {Lin}, \citenamefont {Keimer},\ and\
  \citenamefont {Hinkov}}]{Inosov2010}%
  \BibitemOpen
  \bibfield  {author} {\bibinfo {author} {\bibfnamefont {D.~S.}\ \bibnamefont
  {Inosov}}, \bibinfo {author} {\bibfnamefont {J.~T.}\ \bibnamefont {Park}},
  \bibinfo {author} {\bibfnamefont {P.}~\bibnamefont {Bourges}}, \bibinfo
  {author} {\bibfnamefont {D.~L.}\ \bibnamefont {Sun}}, \bibinfo {author}
  {\bibfnamefont {Y.}~\bibnamefont {Sidis}}, \bibinfo {author} {\bibfnamefont
  {A.}~\bibnamefont {Schneidewind}}, \bibinfo {author} {\bibfnamefont
  {K.}~\bibnamefont {Hradil}}, \bibinfo {author} {\bibfnamefont
  {D.}~\bibnamefont {Haug}}, \bibinfo {author} {\bibfnamefont {C.~T.}\
  \bibnamefont {Lin}}, \bibinfo {author} {\bibfnamefont {B.}~\bibnamefont
  {Keimer}}, \ and\ \bibinfo {author} {\bibfnamefont {V.}~\bibnamefont
  {Hinkov}},\ }\href@noop {} {\bibfield  {journal} {\bibinfo  {journal} {Nature
  Phys.}\ }\textbf {\bibinfo {volume} {6}},\ \bibinfo {pages} {178} (\bibinfo
  {year} {2010})}\BibitemShut {NoStop}%
\bibitem [{\citenamefont {Schr\"oder}\ and\ \citenamefont
  {et~al.}(1998)}]{Schroder1}%
  \BibitemOpen
  \bibfield  {author} {\bibinfo {author} {\bibfnamefont {A.}~\bibnamefont
  {Schr\"oder}}\ and\ \bibinfo {author} {\bibnamefont {et~al.}},\ }\href@noop
  {} {\bibfield  {journal} {\bibinfo  {journal} {Phys. Rev. Lett.}\ }\textbf
  {\bibinfo {volume} {80}},\ \bibinfo {pages} {5623} (\bibinfo {year}
  {1998})}\BibitemShut {NoStop}%
\bibitem [{\citenamefont {Schr\"oder}\ and\ \citenamefont
  {et~al.}(2000)}]{Schroder2}%
  \BibitemOpen
  \bibfield  {author} {\bibinfo {author} {\bibfnamefont {A.}~\bibnamefont
  {Schr\"oder}}\ and\ \bibinfo {author} {\bibnamefont {et~al.}},\ }\href@noop
  {} {\bibfield  {journal} {\bibinfo  {journal} {Nature (London)}\ }\textbf
  {\bibinfo {volume} {407}},\ \bibinfo {pages} {351} (\bibinfo {year}
  {2000})}\BibitemShut {NoStop}%
\bibitem [{\citenamefont {Varma}\ \emph {et~al.}(2015)\citenamefont {Varma},
  \citenamefont {Zhu},\ and\ \citenamefont {Schr\"oder}}]{SchroderZhuV2015}%
  \BibitemOpen
  \bibfield  {author} {\bibinfo {author} {\bibfnamefont {C.~M.}\ \bibnamefont
  {Varma}}, \bibinfo {author} {\bibfnamefont {L.}~\bibnamefont {Zhu}}, \ and\
  \bibinfo {author} {\bibfnamefont {A.}~\bibnamefont {Schr\"oder}},\ }\href
  {\doibase 10.1103/PhysRevB.92.155150} {\bibfield  {journal} {\bibinfo
  {journal} {Phys. Rev. B}\ }\textbf {\bibinfo {volume} {92}},\ \bibinfo
  {pages} {155150} (\bibinfo {year} {2015})}\BibitemShut {NoStop}%
\bibitem [{\citenamefont {Bultinck}\ \emph {et~al.}(2020)\citenamefont
  {Bultinck}, \citenamefont {Khalaf}, \citenamefont {Liu}, \citenamefont
  {Chatterjee}, \citenamefont {Vishwanath},\ and\ \citenamefont
  {Zaletel}}]{Zaletel2020}%
  \BibitemOpen
  \bibfield  {author} {\bibinfo {author} {\bibfnamefont {N.}~\bibnamefont
  {Bultinck}}, \bibinfo {author} {\bibfnamefont {E.}~\bibnamefont {Khalaf}},
  \bibinfo {author} {\bibfnamefont {S.}~\bibnamefont {Liu}}, \bibinfo {author}
  {\bibfnamefont {S.}~\bibnamefont {Chatterjee}}, \bibinfo {author}
  {\bibfnamefont {A.}~\bibnamefont {Vishwanath}}, \ and\ \bibinfo {author}
  {\bibfnamefont {M.~P.}\ \bibnamefont {Zaletel}},\ }\href {\doibase
  10.1103/PhysRevX.10.031034} {\bibfield  {journal} {\bibinfo  {journal} {Phys.
  Rev. X}\ }\textbf {\bibinfo {volume} {10}},\ \bibinfo {pages} {031034}
  (\bibinfo {year} {2020})}\BibitemShut {NoStop}%
\bibitem [{\citenamefont {Hofmann}\ \emph {et~al.}(2021)\citenamefont
  {Hofmann}, \citenamefont {Khalaf}, \citenamefont {Vishwanath}, \citenamefont
  {Berg},\ and\ \citenamefont {Lee}}]{Berg2021}%
  \BibitemOpen
  \bibfield  {author} {\bibinfo {author} {\bibfnamefont {J.~S.}\ \bibnamefont
  {Hofmann}}, \bibinfo {author} {\bibfnamefont {E.}~\bibnamefont {Khalaf}},
  \bibinfo {author} {\bibfnamefont {A.}~\bibnamefont {Vishwanath}}, \bibinfo
  {author} {\bibfnamefont {E.}~\bibnamefont {Berg}}, \ and\ \bibinfo {author}
  {\bibfnamefont {J.~Y.}\ \bibnamefont {Lee}},\ }\href@noop {} {\  (\bibinfo
  {year} {2021})},\ \Eprint {http://arxiv.org/abs/Preprint available at
  http://arxiv.org/abs/2105.12112} {arXiv:Preprint available at
  http://arxiv.org/abs/2105.12112} \BibitemShut {NoStop}%
\bibitem [{\citenamefont {Weber}\ \emph {et~al.}(2014)\citenamefont {Weber},
  \citenamefont {Giamarchi},\ and\ \citenamefont {Varma}}]{Weber-Giam-V}%
  \BibitemOpen
  \bibfield  {author} {\bibinfo {author} {\bibfnamefont {C.}~\bibnamefont
  {Weber}}, \bibinfo {author} {\bibfnamefont {T.}~\bibnamefont {Giamarchi}}, \
  and\ \bibinfo {author} {\bibfnamefont {C.~M.}\ \bibnamefont {Varma}},\ }\href
  {\doibase 10.1103/PhysRevLett.112.117001} {\bibfield  {journal} {\bibinfo
  {journal} {Phys. Rev. Lett.}\ }\textbf {\bibinfo {volume} {112}},\ \bibinfo
  {pages} {117001} (\bibinfo {year} {2014})}\BibitemShut {NoStop}%
\bibitem [{\citenamefont {Nandkishore}\ and\ \citenamefont
  {Levitov}(2011)}]{Nand_levitov_BG}%
  \BibitemOpen
  \bibfield  {author} {\bibinfo {author} {\bibfnamefont {R.}~\bibnamefont
  {Nandkishore}}\ and\ \bibinfo {author} {\bibfnamefont {L.}~\bibnamefont
  {Levitov}},\ }\href {\doibase 10.1103/PhysRevLett.107.097402} {\bibfield
  {journal} {\bibinfo  {journal} {Phys. Rev. Lett.}\ }\textbf {\bibinfo
  {volume} {107}},\ \bibinfo {pages} {097402} (\bibinfo {year}
  {2011})}\BibitemShut {NoStop}%
\bibitem [{\citenamefont {Zhu}\ \emph {et~al.}(2013)\citenamefont {Zhu},
  \citenamefont {Aji},\ and\ \citenamefont {Varma}}]{Lijun-graphene}%
  \BibitemOpen
  \bibfield  {author} {\bibinfo {author} {\bibfnamefont {L.}~\bibnamefont
  {Zhu}}, \bibinfo {author} {\bibfnamefont {V.}~\bibnamefont {Aji}}, \ and\
  \bibinfo {author} {\bibfnamefont {C.~M.}\ \bibnamefont {Varma}},\ }\href
  {\doibase 10.1103/PhysRevB.87.035427} {\bibfield  {journal} {\bibinfo
  {journal} {Phys. Rev. B}\ }\textbf {\bibinfo {volume} {87}},\ \bibinfo
  {pages} {035427} (\bibinfo {year} {2013})}\BibitemShut {NoStop}%
\bibitem [{\citenamefont {Bok}\ \emph {et~al.}(2016)\citenamefont {Bok},
  \citenamefont {Bae}, \citenamefont {Choi}, \citenamefont {Varma},
  \citenamefont {Zhang}, \citenamefont {He}, \citenamefont {Zhang},
  \citenamefont {Yu},\ and\ \citenamefont {Zhou}}]{Bok_ScienceADV}%
  \BibitemOpen
  \bibfield  {author} {\bibinfo {author} {\bibfnamefont {J.~M.}\ \bibnamefont
  {Bok}}, \bibinfo {author} {\bibfnamefont {J.~J.}\ \bibnamefont {Bae}},
  \bibinfo {author} {\bibfnamefont {H.-Y.}\ \bibnamefont {Choi}}, \bibinfo
  {author} {\bibfnamefont {C.~M.}\ \bibnamefont {Varma}}, \bibinfo {author}
  {\bibfnamefont {W.}~\bibnamefont {Zhang}}, \bibinfo {author} {\bibfnamefont
  {J.}~\bibnamefont {He}}, \bibinfo {author} {\bibfnamefont {Y.}~\bibnamefont
  {Zhang}}, \bibinfo {author} {\bibfnamefont {L.}~\bibnamefont {Yu}}, \ and\
  \bibinfo {author} {\bibfnamefont {X.~J.}\ \bibnamefont {Zhou}},\ }\href
  {\doibase 10.1126/sciadv.1501329} {\bibfield  {journal} {\bibinfo  {journal}
  {Science Advances}\ }\textbf {\bibinfo {volume} {2}} (\bibinfo {year}
  {2016}),\ 10.1126/sciadv.1501329},\ \Eprint
  {http://arxiv.org/abs/https://advances.sciencemag.org/content/2/3/e1501329.full.pdf}
  {https://advances.sciencemag.org/content/2/3/e1501329.full.pdf} \BibitemShut
  {NoStop}%
\end{thebibliography}
%\end{document}


\end{document}